\documentclass[aps,prl,twocolumn,nofootinbib,longbibliography,superscriptaddress]{revtex4-2}
\usepackage{amsfonts}
\usepackage{amscd}
\usepackage{amsmath}
\usepackage{amssymb}
\usepackage{newtxtext,newtxmath}
\usepackage{here}
\usepackage[dvipdfmx]{graphicx}
\usepackage{dcolumn}
\usepackage{bm}
\usepackage{xcolor}
\usepackage[%
  colorlinks=true,
  urlcolor=blue,
  linkcolor=blue,
  citecolor=blue
]{hyperref}
\usepackage{tabularx}
\usepackage{etoolbox}
\usepackage{braket}
\usepackage[caption=false]{subfig}
\newcommand{\nn}{\nonumber}
\newcommand{\beq}{\begin{equation}}
\newcommand{\eeq}{\end{equation}}

\makeatletter
\let\cat@comma@active\@empty
\makeatother

\allowdisplaybreaks

\usepackage[normalem]{ulem}

\renewcommand\sout{\bgroup \color{blue} \ULdepth=-.5ex \ULset}

\usepackage{hyperref}
\begin{document}
\preprint{aps/}

\title{
Imprinting spiral Higgs waves onto superconductors with vortex beams
}

\author{Takeshi Mizushima}
\email{mizushima@mp.es.osaka-u.ac.jp}
\affiliation{Department of Materials Engineering Science, Osaka University, Toyonaka, Osaka 560-8531, Japan}
\author{Masahiro Sato}
\email{sato.phys@chiba-u.jp}
\affiliation{Department of Physics, Chiba University, Chiba 263-8522, Japan}
\date{\today}

\begin{abstract}
A vortex beam, akin to a quantized vortex in superfluids, possesses inherent orbital angular momentum (OAM), resulting in the propagation of a spiral-shaped wavefront. Here we demonstrate that a pulsed vortex beam with OAM in the terahertz frequency band can induce a spiral Higgs wave, which is a spiral-shaped oscillation mode of the superconducting order parameter. By utilizing the gauge-invariant theory for the superconducting order, we demonstrate that the phase mode is driven to screen the longitudinal magnetic field of the vortex beam, which facilitates the imprinting of the spiral-shaped wavefront and the transfer of OAM to the condensate. Furthermore, we find that increasing the OAM of light amplifies the intensity of the third harmonic generation. These findings highlight the potential of terahertz vortex beams as a spectroscopic probe of collective modes.
\end{abstract}

\maketitle

{\it Introduction.---}
In 1992, Allen {\it et al}.~\cite{all92} discovered the vortex beam, a light beam with a spiral-shaped wavefront around its propagation axis. Similar to a quantum vortex in superfluids, a vortex beam has a phase singularity and carries both orbital angular momentum (OAM) of $m\hbar$ and spin angular momentum (SAM). SAM designates the handedness of light, while the topological charge of OAM, $m\in\mathbb{Z}$, counts the number of phase windings in a single wavelength. The phase singularity gives rise to a doughnut-shaped intensity distribution~\cite{all92,fra17,she19}. Vortex beams have been developed as essential optical techniques for a wide range of optical and physical phenomena, including optical trapping and manipulation~\cite{he95,kug97,nei02,mac02,cur03,dho11}, quantum communications~\cite{mai01,wan12,boz13}, chiral nanostructure fabrication~\cite{oma10,toy12,toy13}, optical vortex knots~\cite{den10}, and astrophysics~\cite{har03}. Although the research field of vortex beams is continuously expanding, the applications for solid state physics are still limited.

The recent development of generating vortex beams with high intensity in the terahertz (THz) band has exciting implications for solid-state physics~\cite{miy14,ari20,sir19,wan20,sir21,ish23,yav23}. THz vortex beams offer a unique opportunity for ultrafast manipulation of states of matter since photon energy is comparable to that of collective excitations in materials, such as spin waves. The coupling between THz vortex beams and ordered spin structure in magnets leads to the OAM dichroic effect~\cite{sir19,sir21} and Faraday effect~\cite{yav23}. Furthermore, the imprint of the helical structure of vortex beams on spin texture has been demonstrated in a semiconductor quantum well~\cite{ish23}. Theoretical studies have also revealed that vortex beams cause nonequilibrium phenomena in magnets, including the generation of spiral spin waves and topological defects~\cite{fuj17,fuj17-2}.

In this paper, we theoretically study the nonlinear optical responses of superconductors to THz vortex beams. These beams align with the distinctive length scale of superconductors, much like long-wavelength spin waves and topological defects in magnets~\cite{fuj17,fuj17-2}. Given that the coherence length of Cooper pairs is extensive, the macroscopic wave function can effectively capture characteristic wavefronts of vortex beams. The THz band is necessary since its photon energy is compatible with the Higgs mode.
Throughout numerical simulations on the effective action of the superconducting order and gauge fields, we demonstrate that pulsed vortex beams significantly enhance the intensity of third harmonic generation (THG) by driving the spiral-shaped Higgs waves in the condensate. This occurs due to the quadratic coupling of the amplitude oscillation to gauge-invariant potentials. The spacetime phase (plasma) fluctuations are responsible for imprinting the spiral-shaped wavefronts to the condensate and amplifying the THG intensity with the OAM of light. These findings suggest that THz vortex beams serve  as spectroscopic probes for collective modes.

{\it Laguerre-Gaussian vortex beam.---}
Vortex beams are a class of solutions of Maxwell's equations within the paraxial approximation that carries OAM~\cite{all92,hall}. They are often referred to as Laguerre-Gaussian modes. When considering a monochromatic incident electric field and propagation in the $z$ direction, the field configuration can be expressed as 
${\bm E}^{\rm ext}({\bm x},t)= i\Omega[\hat{\bm e}_{s}u_{p,m}+\hat{\bm z}\frac{i}{k}(\hat{\bm e}_{s}\cdot{\bm \nabla})u_{p,m}]e^{-i\Omega t}$, where $\hat{\bm e}_{s} = (\hat{\bm x} + i s\hat{\bm y})/\sqrt{2}$ is the polarization vector for circularly polarized light and $s=\pm 1$ is the SAM of light. Let ${\bm x}=(\rho,\theta,z)$ be the cylindrical coordinate, where $\rho=\sqrt{x^2+y^2}$ and $\theta={\rm arctan}(y/x)$. The spatial profile of $u_{p,m}$ at the focal plane ($z=0$) is given by 
\beq
u_{p,m}(\rho,\theta,z=0) =
\left(
\frac{\rho}{w_0} 
\right)^{|m|}
e^{-\rho^2/w^2_0}e^{im\theta}
L^{|m|}_p\left( \frac{2\rho^2}{w^2_0}\right),
\eeq
where $L^{|m|}_p$ is the associated Laguerre polynomial and $w_0$ is the diameter of the Gaussian beam. Apart from $s=\pm 1$, the Laguerre-Gaussian mode is identified by $p,m\in \mathbb{Z}$. The integer $p$ corresponds to the number of nodes for the radial direction, while $m$ represents the topological charge associated with the OAM of light. This delineates the phase accumulated in multiples of $2\pi$ when traveling around the mode circumference. The winding number of the spiral-shaped wavefront can determine the helicity or handedness of the wavefront. Vortex beams with $m\neq 0$ have a ``doughnut-shaped'' intensity profile, with peak intensity around $\rho \sim w_0$. Gaussian beams, correspond to $m=0$, have been employed in the spectroscopy of the Higgs excitations~\cite{mat13,mat14,she15,kat18,nak19,nak20,kat20,shi20,chu20,vas21,gra22}.
Unlike conventional Gaussian beams, the vortex beam involves a longitudinal component of the magnetic field, ${B}^{\rm ext}_z = ({\bm \nabla}\times {\bm A}^{\rm ext})_z\propto\cos(J\theta - \Omega t)$. 
Below we show that in superconductors, the phase mode screens the longitudinal field, which is essential for imprinting the spiral wavefront of light to the condensate.

{\it Nonlinear dynamics induced by vortex beams.---}
To study the nonlinear dynamics of superconductors induced by vortex beams, we start with the effective action at $T=0$~\cite{stoofPRB93,varma,vorontsovPRB16}
\begin{align}
\mathcal{S} = \int dx\bigg[&
-\tau \left|\mathcal{D}_{t}\Delta(x)\right|^2
- \gamma \Delta^{\ast}(x)\mathcal{D}_{t}\Delta(x)
+\kappa \left|\bm{\mathcal{D}}\Delta(x)\right|^2 \nn\\
&+ \alpha |\Delta(x)|^2+\frac{\beta}{2} |\Delta(x)|^4
+ \frac{{\bm B}^2(x)-{\bm E}^2(x)}{8\pi}
\bigg],
\label{eq:Seff}
\end{align}
where $\Delta(x)$ is the superconducting order parameter. Here we introduce $x\equiv({\bm x},t)$, and $(\tau,\kappa,\alpha)$ are microscopically determined~\cite{SM}. The wave speed and coherence length are defined as $v^2\equiv\kappa/\tau=v^2_{\rm F}/3$ and $\xi^2_0\equiv\kappa/|\alpha|=v^2/2\Delta^2_0$, respectively, where $\Delta_0$ is the superconducting gap at equilibrium and $v_{\rm F}$ is the Fermi velocity of normal electrons. These values characterize the wave propagation and length scale of $\Delta(x)$. The scalar and vector potentials, $(\Phi,\mathcal{\bm A})$, are coupled to $\Delta$ through covariant derivatives, $\mathcal{D}_{t} \equiv \partial _{t} + 2ie\Phi$ and 
$\bm{\mathcal{D}} \equiv {\bm \nabla}-i\frac{2e}{c}\bm{\mathcal A}$. The internal electromagnetic (EM) fields in the superconductor are defined as ${\bm B}={\bm \nabla}\times {\bm A}$ and ${\bm E} = - {\bm \nabla}\phi - \frac{1}{c} \partial_t {\bm A}$, where $\phi\equiv \Phi-\phi^{\rm ext}$ and ${\bm A}\equiv \bm{\mathcal A}-{\bm A}^{\rm ext}$ are the scalar and vector potentials in the superconductor, respectively, and the external potentials generated by vortex beams are represented by $\phi^{\rm ext}=0$ and ${\bm A}^{\rm ext}={\bm E}^{\rm ext}/i\Omega$. In this paper, we set $\hbar=1$. The first term in Eq.~\eqref{eq:Seff} pertains to collective excitations, while the second term, which involves $\gamma$, contributes to the dissipation of these modes back to equilibrium. When the energy scale of the excitations satisfies $\omega \lesssim 2\Delta_0$, the damping is suppressed because of the absence of the fermionic excitations. Therefore, we determine $\gamma$ to satisfy $\gamma \ll \tau\omega$.

\begin{figure}[t!]
\includegraphics[width=85mm]{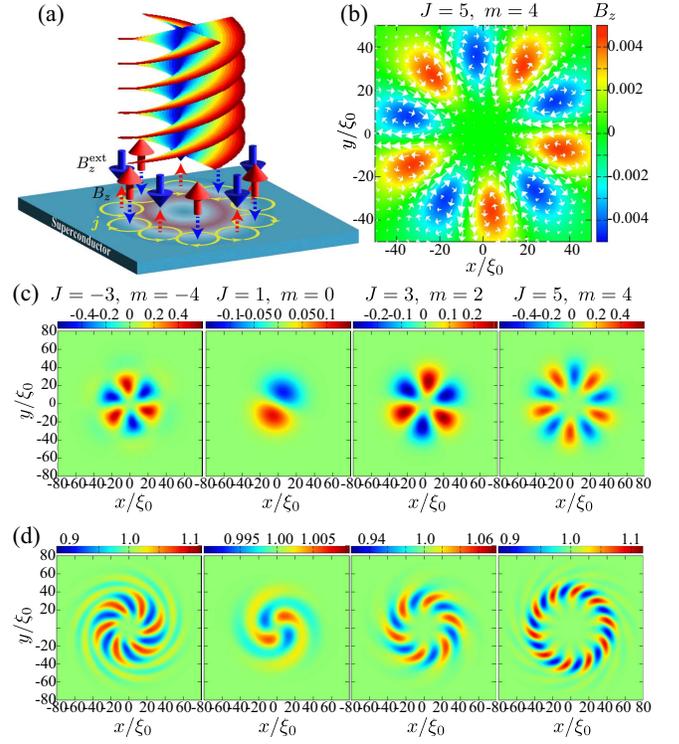}
\caption{(a) Schematics of our setup, where the spiral shape depicts the wavefront of the vortex beam with $m=4$. The vortex beam with $m\neq 0$ involves $B_z^{\rm ext}$, which is screened by the internal field $B_z$ and the supercurrent density ${\bm j}$. (b) Spatial profiles of $B_z({\bm x},t)$ (color map) and ${\bm j}({\bm x},t)$ (arrows) at $t=51t_{\Delta}$, driven by the vortex beam with $(m,s)=(4,1)$. (c,d) Snapshots of the charge density $\rho({\bm x},t)$ at $t=51t_{\Delta}$ (c) and the condensate amplitude $|\Delta({\bm x},t)|$ at $t=153t_{\Delta}$ (d), induced by vortex beams with $J=m+s$, where $t_{\Delta}\equiv \hbar/\Delta_0= O(1~{\rm ps})$ represents the time scale of the condensate. The intensity of the pulsed vortex beam almost becomes maximum around $t=62t_{\Delta}$. In all data, we set $\Omega=1.025\Delta_0$, close to the nonlinear Higgs resonance.}
\label{fig:snap}
\end{figure}

The equation of motion (EOM) for $\Delta$, derived from Eq.~\eqref{eq:Seff}, has a wave-like form with a relaxation term~\cite{SM}. The gauge fields are governed by Amp\`{e}re's law and Gauss's law,
\begin{align}
{\bm \nabla}\times {\bm B}(x)  -\frac{1}{c}\partial_t {\bm E}(x) = \frac{4\pi}{c}{\bm j}(x), \;
{\bm \nabla}\cdot {\bm E}(x)  = 4\pi \rho(x),
\end{align} 
where the charge and supercurrent densities are 
$\rho (x) = 2ie\tau \Delta^{\ast}(x) \mathcal{D} _t \Delta (x) + {\rm c.c.}$ and 
$j_{\mu} (x) = -2ei\kappa
\Delta^{\ast}(x)\mathcal{D}_{\mu}\Delta(x)+{\rm c.c.}$,
respectively. We introduce the dissipation of the EM fields by replacing ${\bm j}$ with ${\bm j}_{\rm tot}={\bm j}+{\bm j}_{\rm n}$. Here ${\bm j}_{\rm n}= \sigma _{\rm n}{\bm E}$ is the dissipative current that follows the Ohmic law. 
The EOMs obey the charge conservation law, $\partial _t\rho + {\bm \nabla}\cdot{\bm j}=0$. 

Equation~\eqref{eq:Seff} is the nonrelativistic ${\rm U}(1)$ Higgs model, which describes the collective modes in bulk superconductors~\cite{varma}. 
Let us consider small amplitude and phase fluctuations of the condensate around $\Delta_0$, 
$\Delta(x) =[ \Delta_0 + \delta \Delta(x)]e^{i\varphi(x)}$. For $|\Delta(x)|=\Delta_0$, the ${\rm U}(1)$ phase ($\varphi(x)$) is absorbed into the gauge fields through the gauge transformation. As a result, the longitudinal component of the gauge field gains a mass gap of $\omega_{\rm p}= \sqrt{4\pi e^2 n/m_{\rm e}}$, where $n$ is the electron density and $m_{\rm e}$ is the mass of electrons. Then, the only low-lying collective excitation is the Higgs mode with the dispersion, $\omega^2_{\rm H}({\bm q}) = {4\Delta_0^2+v^2q^2}$. The action has been utilized for the numerical simulations of spontaneous vortex formation~\cite{don07} and collisional dynamics of vortices~\cite{mor88}. For numerical calculations, we take the temporal gauge that $\phi=0$ in all time~\cite{mor88,kat91,du94,gro96}, and implement methods from the Hamiltonian formalism of the lattice gauge theory to ensure gauge invariance and charge conservation~\cite{kog79,mor88,SM}. 

We first consider a superconducting film with a thickness ($d$) that is less than the London penetration depth $\lambda$ and $\xi_0$. Then, the dynamics of $\Delta(x)$ and ${\bm j}(x)$ can be restricted to two dimensions. 
Starting the equilibrium state, $\Delta({\bm x},t=0)=\Delta_0$, we irradiate a pulsed vortex beam for $t>0$ [Fig.~\ref{fig:snap}(a)]. Since the wavelength of THz vortex beams is significantly longer than both $d$ and $\lambda$, we can disregard the $z$-dependence of ${\bm A}^{\rm ext}$. The vector potential for the pulsed vortex beam with $(m,s)$ is given by
\beq
{\bm A}^{\rm ext}({\bm x},t) = {\rm Re}\left\{\frac{A_0u_{p,m}({\bm x})}{\max | u_{p,m}({\bm x})|}
\exp\left[
-\left( \frac{t-t_{0}}{\sigma}\right)^2 -i\Omega t
\right]\hat{\bm e}_s\right\},
\label{eq:pulse}
\eeq
where $\Omega$ and $\sigma\equiv 2\pi n_{\rm p}/\Omega$ are the frequency and the full width at half-maximum of the beam intensity, respectively. In this work, we consider vortex beams with $p=0$.
In the following calculations, we take the number of cycles of the pulse field and the beam waist as $n_{\rm p}=5$ and $w_0=20\xi_0$, respectively. We also set $t_0=2\sigma$ and the size of the $xy$ plane to $x,y\in[-200\xi_0,200\xi_0]$. In Ref.~\cite{SM}, we clarify that $n_{\rm p}$ and $w_0$ do not alter the dynamics of $\Delta(x)$ and EM waves. In conventional superconductors, $\xi_0$ and $\lambda$ are $O(10-100~{\rm nm})$. For instance, bulk Nb has $\xi_0\sim\lambda\sim 40~{\rm nm}$~\cite{mes69} and NbN thin films have a large value of the Ginzburg-Landau (GL) parameter $\kappa_{\rm GL}\equiv \lambda/\xi_0$ with $\xi_0\sim 5~{\rm nm}$ and $\lambda\sim 200~{\rm nm}$~\cite{cho08}. The time scale is represented by $t_{\Delta}\equiv\hbar/\Delta_0=O(1~{\rm ps})$ with $\Delta_0=O(1~{\rm meV})$. In the following calculations, the coefficient in Eq.~\eqref{eq:pulse} is fixed to $2|e|\xi_0A_0/\hbar c=0.4$, corresponding to the maximum electric field of $E_0\equiv \Omega A_0 =0.4~{\rm kV}/{\rm cm}$ with $\xi_0=10~{\rm nm}$ and $\Omega = 0.5~{\rm THz}$. In the current work, the heating effect is not taken into account since laser absorption has minimal impact on THz vortex beams. For ultraviolet-visible vortex beams, however, the heating effect becomes more substantial~\cite{tod18,tod23,fuj17-2}.

Figure~\ref{fig:snap}(b) shows the spatial profile of the internal magnetic field $B_z({\bm x},t)$ and the supercurrent density ${\bm j}({\bm x},t)$ (arrows) at $t=51t_{\Delta}\sim t_0$, driven by the circularly polarized vortex beam with $(m,s)=(4,1)$. Here we set the GL parameter to $\kappa_{\rm GL}\equiv \lambda/\xi_0 = 10$. The frequency of light is set to be close to the nonlinear resonance of the Higgs mode, $\Omega = 1.025\Delta_0$. 
The induced supercurrent generates the internal field $B_z({\bm x},t)\propto - B^{\rm ext}_z({\bm x},t)$, which screens the external field $B^{\rm ext}_z({\bm x},t)$. 
The charge conservation dictates that the inhomogeneous current density in the plane is accompanied by a fluctuation in the charge density as $\partial _t \rho = - {\bm \nabla}\cdot{\bm j}$. In Fig.~\ref{fig:snap}(c), we present the snapshots of $\rho({\bm x},t)$ at $t=51t_{\Delta}$ for $m=-4$, $0$, $2$, and $4$ ($J=-3$, $1$, $3$, and $5$), which yield the oscillation pattern that reflects the total angular momentum of light, $J$. In Fig.~\ref{fig:snap}(d), we observe that the oscillation pattern of the condensate amplitude along the azimuthal direction carries twice the total angular momentum $2J$.

\begin{figure}[t!]
\includegraphics[width=85mm]{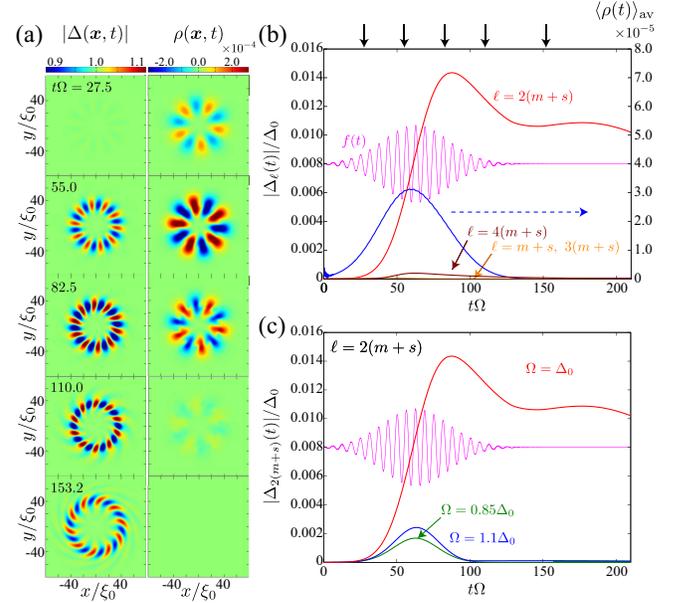}
\caption{(a) Snapshots of $|\Delta({\bm x},t)|$ and $\rho({\bm x},t)$ at $t\Omega=27.5$, $55.0$, $82.5$, $110.0$, and $137.5$ for $\Omega = 1.025\Delta_0$ and $(m,s)=(4,1)$, where $\Omega^{-1}=0.6~{\rm ps}$ for $\Delta_0=1~{\rm meV}$. (b) Time evolutions of the projections of $\Delta$ onto angular momentum eigenstates ($\Delta_{\ell}$) with $\ell=J$, $2J$, $3J$, and $4J$. The amplitudes of the other $\ell$ components are negligibly small. In (a), we also plot the spatially averaged charge density fluctuation, $\langle |\rho(t)|\rangle _{\rm av}$. (c) Time evolutions of $\Delta_{\ell=2J}$ for $\Omega/\Delta_0=0.85$, $1.0$, and $1.1$. In (b) and (c), the magenta curve shows the temporal profile of the applied field $f(t)=\cos(\Omega t)e^{-(t-t_0)^2/\sigma^2}$. }
\label{fig:OAM}
\end{figure}

Figure~\ref{fig:OAM} shows how vortex beams induce the dynamics of the condensate and charge density. Here we focus on $(m,s)=(4,1)$ and $\Omega = 1.025\Delta_0$. In Fig.~\ref{fig:OAM}(a), we display the snapshots of $|\Delta({\bm x},t)|$ and $\rho({\bm x},t)$ after irradiating the pulsed wave in Eq.~\eqref{eq:pulse} onto the condensate. For $t\lesssim t_0/2=10\pi t_{\Delta}$, the vortex beam mainly stimulates the charge density with the peak intensity around $t\sim t_0$. Subsequently, a spiral-shaped wave appears on the condensate $\Delta(x)$, persisting even after the charge density fluctuation $\rho(x)$ is fully screened. We expand the condensate wave function in terms of the eigenstates of the angular momentum as $\Delta(\theta,t) \equiv \int \Delta({\bm x},t)d\rho = \sum_{\ell}e^{i\ell\theta}\Delta_{\ell}(t)$ ($\ell\in\mathbb{Z}$). In Fig.~\ref{fig:OAM}(b), we find that $\Delta_{\ell}$ with $\ell = 2J$ significantly increases around $t\sim t_0$, while the other components with $\ell \neq 2J$ are almost negligible. The time evolution of the net charge density excitation, $\langle \rho(t)\rangle _{\rm av}\equiv \int d{\bm x}|\rho({\bm x},t)|$, is also shown in Fig.~\ref{fig:OAM}(b). This excitation follows the envelope of the pulsed field $e^{-(t-t_0)^2/\sigma^2}$. The charge density is excited before the oscillation of the condensate amplitude grows, and it returns to neutral once the pulse is off. As mentioned below, the charge density oscillation is associated with the off-resonant phase excitation and only transiently excited. 
In Fig.~\ref{fig:OAM}(c), we observe a sharp reduction in the intensity of $\Delta_{2J}(t)$ as the frequency $\Omega$ deviates from the nonlinear resonance to the Higgs mode, $\Omega = \Delta_0$. 

{\it Gauge-invariant response and enhancement of THG.---}
Let us now examine the results in Figs.~\ref{fig:snap} and \ref{fig:OAM}, based on the analysis of the effective action. The phase of the superconducting order, $\varphi$, defined by $\Delta=|(\Delta_0+\delta\Delta)|e^{i\varphi}$, is essential for the gauge invariance of the theory. Here we clarify that vortex beams carrying nonzero OAM stimulate phase excitations, which strengthen the gauge-invariant quadratic coupling of the vortex beams to the Higgs mode ($\delta\Delta$) and amplifies the nonlinear current responses. 

To uncover the role of the phase fluctuation, let us focus on the gauge-invariant potential, $\tilde{\bm{\mathcal{A}}}(x)\equiv \bm{\mathcal A}(x) -\frac{c}{2e}{\bm \nabla}\varphi(x)$. 
Solving the continuity equation and the Maxwell-Amp\`{e}re law in the linear response regime, the potential reads~\cite{SM}
\beq
\tilde{\bm{\mathcal{A}}}(x) \approx 
-\frac{(\kappa_{\rm GL}/C)^2}{1-(\kappa_{\rm GL}/C)^2}{\bm A}_{\perp}^{\rm ext}(x)
-\frac{ic}{2e}
\frac{{\bm \nabla}[{\bm \nabla}\cdot {\bm A}_{\perp}^{\rm ext}(x)]}{(\omega_{\rm p}/v)^2},
\label{eq:gauge}
\eeq
for $\Omega \sim \Delta_0$, where $C\sim c/v$ and ${\bm A}_{\perp}^{\rm ext}=(A_x^{\rm ext},A_y^{\rm ext},0)$. The first term pertains to the screening effect, while the second term is attributed to the off-resonant phase excitation, ${\bm \nabla}\varphi$. In the low $\kappa_{\rm GL}$ regime, the internal field is induced to screen ${\bm A}^{\rm ext}_{\perp}$ and the first term vanishes at the type-I limit, $\kappa_{\rm GL}\rightarrow 0$. As a result, $B_z^{\rm ext}$ generated by vortex beams is screened by the internal field $B_z$, while the transverse fields $(B^{\rm ext}_x,B^{\rm ext}_y)$ remain unscreened in a thin film. On the other hand, at the type-II limit ($\kappa_{\rm GL}\rightarrow \infty$), the screening current is negligible, and the first term of Eq.~\eqref{eq:gauge} reduces to ${\bm A}^{\rm ext}_{\perp}$.
The second term represents the off-resonant phase excitation, where the THz range is far from the resonance  $\Omega \ll \sqrt{(vq)^2+\omega^2_{\rm p}}\approx \omega_{\rm p}$. The phase excitation is linearly driven by vortex beams as $\varphi(x) \propto {\nabla}\cdot{\bm A}^{\rm ext}_{\perp}(x)\propto \hat{\bm e}_s\cdot{\bm \nabla}e^{im\theta-i\Omega t}u_{p,m}({\bm x})$, indicating that the oscillation pattern of the phase mode reflects the OAM of light as $\varphi(x)\propto m\cos(J\theta-\Omega t)$. This explains the characteristic oscillation pattern of $\rho({\bm x},t)\propto \partial _t \varphi({\bm x},t)\propto m\cos(J\theta-\Omega t)$ shown in Fig.~\ref{fig:snap}(c). Thus the OAM of light can be encoded to the superconductor in the form of the charge density oscillation. It is also worth mentioning that the intensity of the first term in Eq.~\eqref{eq:gauge} is insensitive to the OAM of light, $m$, while the second term is proportional to $m$. In the low $\kappa_{\rm GL}$ regime, the first term in Eq.~\eqref{eq:gauge} is screened, while the impact of the second term becomes prominent. Then, the intensity of the gauge-invariant potential ($\tilde{\bm{\mathcal{A}}}$) amplifies through the off-resonant phase excitation as $m$ increases. This is a unique feature of vortex beams and absent in conventional Gaussian beams with $m=0$.

\begin{figure}[t!]
\includegraphics[width=85mm]{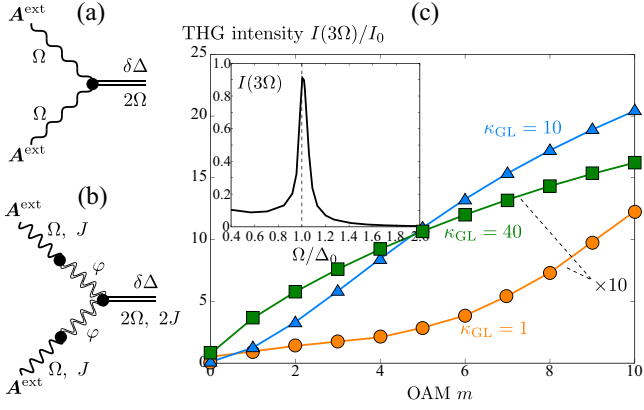}
\caption{(a,b) Diagrams of nonlinear Higgs excitations by vortex beams: (a) The coupling to $({\bm A}^{\rm ext})^2$ and (b) the quadratic coupling through the phase modes. (c) $m$-dependence of the THG intensity $I(3\Omega)/I_0$ at $\Omega = 1.025\Delta_0$ for  $\kappa_{\rm GL}=1$ (circles), $10$ (triangles), and $40$ (squares), where $I_0\equiv c^2/(16\pi e)$. The inset shows the THG intensity, $I(3\Omega)$, for $(m,s)=(4,1)$ and $\kappa_{\rm GL}=10$.}
\label{fig:fs}
\end{figure}

The phase excitation and the screening effect affect the coupling of light to the Higgs mode the THG intensity. From Eq.~\eqref{eq:Seff}, the third-order current response is given as 
\beq
{\bm j}^{(3)}({\bm x},t) = -\frac{16e^2\kappa\Delta_0}{c}\delta\Delta({\bm x},t)\tilde{\bm{\mathcal{A}}}({\bm x},t),
\label{eq:j3}
\eeq
which is mediated by the Higgs excitations ($\delta\Delta$). The Higgs mode ($\delta\Delta$) is coupled to $\tilde{\bm{\mathcal{A}}}$ through the effective action 
\beq
\mathcal{S}_{\delta\Delta A} \equiv \frac{4e^2\kappa}{c^2} \int dx|\delta\Delta(x)|^2\tilde{\bm{\mathcal{A}}}^2(x) .
\label{eq:couple}
\eeq
As mentioned above, the potential is approximated as $\tilde{\bm{\mathcal{A}}}\sim -\frac{2e}{c}{\bm \nabla}\varphi$ in the low $\kappa_{\rm GL}$ regime, and its quadratic form involves $\varphi^2\propto (\hat{\bm e}_s\cdot{\bm \nabla}e^{im\theta-i\Omega t})^2\propto \sin^2(J\theta-\Omega t)$, which generates the spiral wave in the condensate with a period of $ 2\pi w_0 /2J$.
For Gaussian beams with $m=0$, only the direct coupling of $\delta\Delta$ and $(\tilde{\bm{\mathcal{A}}})^2\approx ({\bm A}^{\rm ext})^2$ in Fig.~\ref{fig:fs}(a) is possible and the phase mode is not active. In the case of vortex beams with $m\neq 0$, however, the inhomogeneity causes the off-resonant phase excitations and enhances the gauge-invariant potential with increasing $|m|$. Figure \ref{fig:fs}(b) depicts the impact of the phase mode on $\tilde{\bm{\mathcal{A}}}$, which strengthens both the nonlinear Higgs excitation and the nonlinear current response.

To verify the aforementioned scenario, we compute the THG intensity for different values of $\Omega$, $m$, and $\kappa_{\rm GL}$. The THG intensity is defined as $I(3\Omega) = \int dt I(t)e^{i3\Omega t}$ with $I(t) \equiv \int d{\bm x}|{\bm j}({\bm x},t)|$. As shown in Eq.~\eqref{eq:j3}, the nonlinear Higgs excitation contributes to the THG as $I(3\Omega)\propto \int d{\bm q} \tilde{\bm{\mathcal{A}}}({\bm q},\Omega)\tilde{\bm{\mathcal{A}}}^2(-{\bm q},-2\Omega)/[(2\Omega)^2-(\omega_{\rm H}({\bm q}))^2]$, and the gauge-invariant potential involves the off-resonant phase excitations. In Fig.~\ref{fig:fs}(c), the inset displays the THG intensity, $I(3\Omega)$, for $(m,s)=(4,1)$. The spectrum exhibits a distinct peak at the nonlinear Higgs resonance frequency $\Omega = \Delta_0$. The main panel of Fig.~\ref{fig:fs}(c) indicates that the weak intensity of the THG at $m=0$ increases as $|m|$ increases. This aligns with the scenario where a vortex beam with $m\neq 0$ linearly drives a spacetime fluctuation of the phase mode, enhancing the nonlinear coupling of the Higgs mode to light through the gauge-invariant potential. 

The THG intensity is influenced by $\kappa_{\rm GL}$. As shown in Fig.~\ref{fig:fs}(c), the THG intensity at $\kappa_{\rm GL}=1$ is approximately ten times weaker than that at $\kappa_{\rm GL}=10$ due to the strong screening effect, but it is significantly amplified by the OAM of light. The $\kappa_{\rm GL}$ dependencies are attributed to the interplay between the screening effect and off-resonant plasma oscillation. As mentioned in Eq.~\eqref{eq:gauge}, the gauge-invariant potential $\tilde{\bm{\mathcal{A}}}$ consists of ${\bm A}^{\rm ext}$ and the off-resonant phase excitation. In the low $\kappa_{\rm GL}$ regime ($\kappa _{\rm GL}=1,10$), the former is screened, and $\tilde{\bm{\mathcal{A}}}$ is dominated by the latter term which amplifies with $m$. For $\kappa_{\rm GL}\gg 1$, however, the screening effect weakens and the potential reduces to $\tilde{\bm{\mathcal{A}}}\approx {\bm A}^{\rm ext}$. The nonlinear current response in $\kappa_{\rm GL}=40$ is governed by ${\bm A}^{\rm ext}$ rather than off-resonant plasma excitations. As a result, the increase of the THG intensity with $m$ is suppressed for $\kappa_{\rm GL} \gg 1$. There are optimal values of $\kappa_{\rm GL}$ that maximize the THG intensity~\cite{SM}. We also note that the result in Fig.~\ref{fig:fs}(c) is insensitive to $w_0$~\cite{SM}.

\begin{figure}[t!]
\includegraphics[width=85mm]{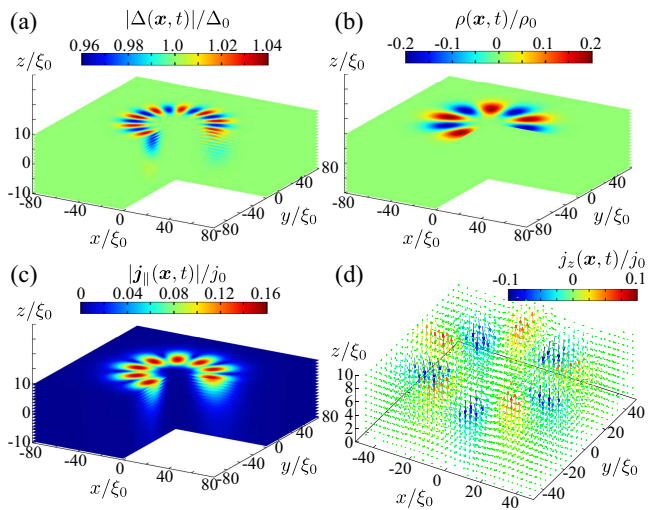}
\caption{Snapshots of $|\Delta({\bm x},t)|$ (a), $\rho({\bm x},t)$ (b), $|{\bm j}_{\parallel}({\bm x},t)|$ (c), and ${\bm j}({\bm x},t)$ (d) at $t=51t_{\Delta}$ after the irradiation of the pulsed vortex beam with $(m,s)=(4,1)$ and $\Omega=1.025\Delta_0$. In all data, we set $\kappa_{\rm GL}=10$.}
\label{fig:3d}
\end{figure}

{\it Spiral waves on the surface of SCs.---}
So far, we have focused on two-dimensions. However, comparable outcomes can be seen even when we take into account a finite thickness along the $z$-axis. 
Consider a superconducting film with a thickness of $d=20\xi_0$ and the penetration depth of $\lambda=10\xi_0$ ($\kappa_{\rm GL}=10$). The vortex beam is irradiated towards the upper surface of the film at $z=d/2$.
Figure~\ref{fig:3d} shows the snapshots of $|\Delta({\bm x},t)|$ (a), $\rho({\bm x},t)$ (b), and ${\bm j}_{\parallel}=(j_x,j_y)$ (c), after the irradiation of the vortex beam with $(m,s)=(4,1)$. Here we set $\Omega=1.025\Delta_0$ and $w_0=20\xi_0$. Figure~\ref{fig:3d}(d) is the vectorial plot of $(j_x,j_y,j_z)$. 
In a similar manner to the results observed in a two-dimensional film, the vortex beam induces spiral waves in both the condensate amplitude [Fig.~\ref{fig:3d}(a)] and the phase excitation [Fig.~\ref{fig:3d}(b)], which reflect twice the total angular momentum $2J$ and the total angular momentum $J$, respectively. The amplitude oscillation penetrates the skin depth within $\lambda$, while alternative positive and negative charge distributions accumulate on the surface as $\rho({\bm x},t)\sim A\sin(J\theta-\Omega t)\delta(z-d/2)$. Such charge distribution creates a three-dimensional flow of the supercurrent density, depicted in Fig.~\ref{fig:3d}(c,d).
Two-dimensional simulations can accurately represent the nonlinear Higgs excitation and THG induced by vortex beams. These characteristics remain consistent as long as the thickness is much shorter than the laser wavelength.

{\it Concluding remarks.---}
In this study, we have investigated the nonlinear optical responses of superconductors to vortex beams and examined the possibility of utilizing OAM as an additional degree of freedom. We have demonstrated that vortex beams with nonzero OAM induce the spiral-shaped oscillations in both the condensate and charge density. The phase modes linearly driven by vortex beams facilitate the imprinting of the spiral-shaped wavefront and the transfer of OAM to the condensate. We have also found that increasing the OAM of light amplifies the THG intensity. In Ref.~\cite{SM}, we have further discussed how light and material parameters impact on our key findings. Here we have identified optimal values for the Ginzburg-Landau parameter and the beam waist to maximize nonlinear Higgs excitation and THG intensity.

We have used the effective action which only takes into account the diamagnetic Higgs response, without considering the contribution of Bogoliubov quasiparticles. For Gaussian beams, however, it has been unveiled that quasiparticle pair excitations play a crucial role in the THG~\cite{cea16,tsu15}. In addition, the third-order paramagnetic current responses~\cite{juj18,sil19} may also come into play for vortex beams with nonzero OAM even in clean superconductors. Further research is needed to calculate the THG from the gauge-invariant microscopic theory~\cite{lut08,ros21,rad22}, including both quasiparticles and the Higgs mode via paramagnetic and diamagnetic channels, beyond Eq.~\eqref{eq:Seff}.

As we mentioned in the introduction, applications of structured light have been gradually extending to condensed matter physics~\cite{sir19,wan20,sir21,ish23,yav23,and06,fuj17,fuj17-2,sim08,che18,fuj18,fuj19,ket21,gun23}. On top of vortex beams, several types of topological light have been developed~\cite{zha09,wu15,he22}. The present study and these works will further accelerate the marriage of topological light with condensed matter physics.

\begin{acknowledgments}
We thank Naoto Tsuji, Takashige Omatsu, and Yuki Kawaguchi for useful discussions.
This work was supported by a Grant-in-Aid for Scientific Research on Innovative Areas ``Quantum Liquid Crystals'' (Grant No.~JP19H05825 and No.~JP22H04480) and ``Evolution of Chiral Materials Science using Helical Light Fields'' (Grants No. JP22H05131, No.~JP23H04576) from JSPS of Japan, and JSPS KAKENHI (Grant No.~JP20K03860, No.~JP20H01857, No.~JP20H01830, No.~JP20H01849, No.~JP21H01039, and No.~JP22H01221).
\end{acknowledgments}

\bibliography{LGbeam}

\end{document}


\preprint{aps/}
\title{Supplemental Materials for ``Imprinting spiral Higgs waves onto superconductors \\ with vortex beams''}

\author{Takeshi Mizushima}
\email{mizushima@mp.es.osaka-u.ac.jp}
\affiliation{Department of Materials Engineering Science, Osaka University, Toyonaka, Osaka 560-8531, Japan}
\author{Masahiro Sato}
\email{sato.phys@chiba-u.jp}
\affiliation{Department of Physics, Chiba University, Chiba 263-8522, Japan}

\date{\today}

\maketitle

In this supplemental material, we present the detail of numerical and analytical calculations on the dynamics of superconductors by vortex beams. {In Sec.~S1,} we begin by presenting the effective theory for the gauge-invariant dynamics of superconducting order, along with numerical techniques based on the Hamiltonian formalism of the lattice gauge theory~\cite{kog79,mor88}. In Secs.~S2 and S3, we examine how the parameters of the superconducting material and the vortex beam influence key findings, such as the spiral-shaped excitations of the superconducting order and the enhancement of the third harmonic generation (THG). We identify optimal values for the Ginzburg-Landau parameter and the beam waist to maximize nonlinear Higgs excitation and THG intensity.

\section{S1. Effective gauge-invariant theory for the dynamics of superconducting order}

\subsection{S1.1 Effective action}

We start with the effective action for the superconducting order $\Delta$ and the electromagnetic (EM) fields~\cite{stoofPRB93,varma,vorontsovPRB16}
\begin{align}
\mathcal{S} = \int dx\bigg[&
-\tau \left|\mathcal{D}_{t}\Delta(x)\right|^2
+\kappa \left|\bm{\mathcal{D}}\Delta(x)\right|^2 \nn \\
&+ \alpha |\Delta(x)|^2+\frac{\beta}{2} |\Delta(x)|^4 
\bigg] 
 + \mathcal{S}_{\rm EM},
\label{eq:Seff}
\end{align}
where the action for the EM fields is given by 
\begin{align}
\mathcal{S}_{\rm EM} = \int dx\frac{{\bm B}^2-{\bm E}^2}{8\pi},
\end{align}
and  the covariant derivatives are defined as $\mathcal{D}_{t} \equiv \partial _{t} + 2ie\Phi$ and 
$\bm{\mathcal{D}} \equiv {\bm \nabla}-i\frac{2e}{c}\bm{\mathcal A}$.
Here we have introduced abbreviation $x\equiv ({\bm x},t)$ and set $\hbar=1$. For the derivation of Eq.~\eqref{eq:Seff}, we assume $T_{\rm c}\ll T_{\rm F}$, where $T_{\rm c}$ and $T_{\rm F}$ are the superconducting transition temperature and the Fermi temperature, respectively.
The EM fields are expressed in terms of scalar and vector potentials $(\phi,{\bm A})$ as ${\bm B}={\bm \nabla}\times {\bm A}$ and ${\bm E} = - {\bm \nabla}\phi - \frac{1}{c} \partial_t {\bm A}$, respectively, and $(\Phi,\bm{\mathcal A})$ are the sum of the internal fields $(\phi,{\bm A})$ and external fields $(\phi^{\rm ext},{\bm A}^{\rm ext})$ 
\beq
\Phi(x) = \phi(x) + \phi^{\rm ext}(x), \quad
\bm{\mathcal{A}}(x) = {\bm A}(x) + {\bm A}^{\rm ext}(x).
\label{eq:Atotal}
\eeq
The external fields, $(\phi^{\rm ext},{\bm A}^{\rm ext})$, are generated by vortex beams.
The values of the coefficients $(\tau,\kappa,\alpha)$ can be determined from the microscopic action at $T\rightarrow 0$, where $\tau = N_{\rm F}/8\Delta_0^2$, $-\alpha =\Delta_0^2 \beta = N_{\rm F}/4$, and $\kappa=N_{\rm F}v^2_{\rm F}/24\Delta^2_0$~\cite{stoofPRB93}. The density of states at the Fermi level, the Fermi velocity of normal electrons, and the superconducting gap at equilibrium are represented by $N_{\rm F}$, $v_{\rm F}$, and $\Delta_0$, respectively. The wave speed and coherence length are defined as $v^2\equiv\kappa/\tau=v^2_{\rm F}/3$ and $\xi^2_0\equiv\kappa/|\alpha|=v^2/2\Delta^2_0$, respectively, and these values characterize the length scale and wave propagation of $\Delta(x)$. The equations of motion for the superconducting order parameter and EM fields are obtained from the extrema of the action $\mathcal{S}$ with respect to $\Delta(x)$, $\phi (x)$, ${\bm A}(x)$ as 
\begin{align}
\tau
\mathcal{D}_t^2
\Delta(x) =& \kappa \bm{\mathcal{D}}^2\Delta (x)
- \beta\left(|\Delta(x) |^2-\Delta^2_0\right)\Delta(x) ,
\label{eq:eom1_app}
\end{align}
and 
\begin{gather}
{\bm \nabla}\times {\bm B}(x)  -\frac{1}{c}\partial_t {\bm E}(x) = \frac{4\pi}{c}{\bm j}(x) , \label{eq:ampere}\\
{\bm \nabla}\cdot {\bm E}(x)  = 4\pi \rho(x) .
\label{eq:eom2_app}
\end{gather} 
The charge density and the supercurrent density are expressed as 
\begin{align}
\rho (x) 
=& -8e^2\tau |\Delta(x)|^2\tilde{\Phi}(x), \\ 
{\bm j} (x) 
=& -\frac{8e^2\kappa}{c} |\Delta(x)|^2 \tilde{\bm{\mathcal A}}(x),\label{eq:current}
\end{align}
respectively, where $\Delta (x)$ is decomposed to the amplitude $|\Delta(x)|$ and the phase $\varphi(x)$. The charge and current densities are related to the gauge-invariant potentials, 
\begin{gather}
\tilde{\Phi}(x)\equiv \Phi(x)+\frac{1}{2e}\partial_t \varphi(x), \\
\tilde{\bm{\mathcal A}}(x) \equiv \bm{\mathcal A}(x)-\frac{c}{2e}{\bm \nabla}\varphi(x).
\end{gather}
The equations of motion for $\Delta(x)$ and the EM fields conserve the electric charge and obey the continuity equation, 
\beq
\partial _t\rho + {\bm \nabla}\cdot{\bm j}=0, \label{eq:cont}
\eeq
which is obtained from $\delta\mathcal{S}/\delta\varphi(x) = 0$.
 
Equation~\eqref{eq:Seff} is the nonrelativistic ${\rm U}(1)$ Higgs model, which properly describes the low-energy dynamics of superconductors coupled to the gauge fields~\cite{varma}. Let us consider small amplitude and phase fluctuations of the condensate around $\Delta_0$, 
$\Delta(x) =[ \Delta_0 + \delta \Delta(x)]e^{i\varphi(x)}$. In the long wavelength limit, the eigenfrequency of the Higgs mode is given as $\omega^2_{\rm H}({\bm q}) = {4\Delta_0^2+v^2q^2}$. When the electric charge is absent, the phase mode has the gapless dispersion, $\omega_{\rm NG}({\bm q})=vq$, which is the Nambu-Goldstone mode associated with the ${\rm U}(1)$ symmetry breaking. 
For three-dimensional bulk superconductors with $|\Delta(x)|=\Delta_0$, the ${\rm U}(1)$ phase, $\varphi(x)$, is absorbed into the gauge fields via the gauge transformation, $A_{\mu}\rightarrow A_{\mu}-(c/2e)\partial_{\mu}\varphi$ and $\phi\rightarrow \phi + \partial_t\varphi/2e$. Consequently, the longitudinal component of the gauge field obtains a mass gap. The dispersions of the EM fields are $\omega^2_{\parallel}({\bm q})=c^2{\bm q}^2+\omega^2_{\rm p}$ for ${\bm q}\perp{\bm A}$ and $\omega^2_{\perp}({\bm q})=v^2{\bm q}^2+\omega^2_{\rm p}$ for ${\bm q}\parallel{\bm A}$. The plasma frequency is defined as $\omega_{\rm p}= \sqrt{4\pi e^2 n/m_{\rm e}}$, where $n=2N_{\rm F}m_{\rm e}v^2_{\rm F}/3$ is the electron density and $m_{\rm e}$ is the mass of the electrons. Therefore, for bulk superconductors, the only low-energy bosonic excitations are the Higgs mode.

We introduce a dissipation term into the effective action to account for the relaxation of the superconducting order and EM fields.
The term with $\gamma > 0$,
\beq
\mathcal{S}_{\rm damp} = -\int  \gamma \Delta^{\ast}(x) \mathcal{D}_t\Delta(x) dx.
\eeq 
causes dissipation back to equilibrium and damping of the collective modes. The damping is suppressed when the energy scale of the excitations is $\omega \lesssim 2\Delta_0$, due to the absence of the fermionic excitations. We set the damping rate $\gamma$ to be much smaller than $\tau\omega $. Additionally, we introduce the dissipation of the EM fields by replacing ${\bm j}$ with ${\bm j}_{\rm tot}={\bm j}+{\bm j}_{\rm n}$. Here ${\bm j}_{\rm n}= \sigma _{\rm n}{\bm E}$ is the dissipative current that follows the Ohmic law.

\subsection{S1.2 Hamiltonian formalism}

For numerical simulations, we implement methods from the Hamiltonian formalism of the lattice gauge theory~\cite{kog79,mor88}, rather than solving the Euler-Lagrange equations. The Hamiltonian formalism has an advantage in that, due to the gauge invariance of the Hamiltonian, Gauss's law in Eq.~\eqref{eq:eom2_app} serves as a constraint that the initial field configuration must satisfy. To model the Hamiltonian, we begin by introducing the momenta $\pi \equiv \partial \mathcal{L}/\partial \dot{\Delta}^{\ast}$ and $\varepsilon_{\mu}\equiv \partial \mathcal{L}/\partial \dot{A}_{\mu}=-E_{\mu}/4\pi c$ as conjugates to the fields $\Delta^{\ast}$ and $A_{\mu}$, where the Lagrangian density is defined as $\mathcal{S}=\int dx\mathcal{L}$. The Hamiltonian density $\mathcal{H}$ is then derived from the Lagrangian using the fields $\Delta,\Delta^{\ast},{\bm A}$ and canonical momenta $\pi,\pi^{\ast},{\bm \varepsilon}=-{\bm E}/4\pi c$ as well as $\phi$,
\begin{align}
\mathcal{H} =& 
-\frac{1}{\tau}|\pi(x)|^2 +\kappa\left|\bm{\mathcal{D}}\Delta(x)\right|^2 
 + \alpha|\Delta(x)|^2 + \frac{\beta}{2}|\Delta(x)|^4\nn \\
&+2ie\Phi(x)\left\{\pi(x)\Delta^{\ast}(x)-\pi^{\ast}(x)\Delta(x)\right\} \nn \\
&+\frac{({\bm \nabla}\times {\bm A}(x))^2}{8\pi} -\frac{{\bm E}^2(x)}{8\pi}+\frac{{\bm E}(x)}{4\pi}\cdot{\bm \nabla}\phi(x).
\end{align}
The Hamiltonian is invariant under the local gauge transformation, $E_{\mu}\rightarrow E_{\mu}$, $A_{\mu}\rightarrow A_{\mu}+\partial_{\mu}\chi$, $\Delta \rightarrow \Delta e^{-i\frac{2e}{c}\chi}$, $\pi \rightarrow \pi e^{-i\frac{2e}{c}\chi}$, and $\phi \rightarrow \phi + \partial_t\chi/c$, where $\chi$ is an arbitrary function. In numerical simulations, we take the temporal (zero electric potential) gauge~\cite{kog79,kat91,du94,gro96}. This means that at all times, $\phi(x)=0$. Essentially, we ensure that $\partial _t \chi(x) =c \phi(x)$, which removes the scalar potential from the equations of motion. The ${\rm U}(1)$ gauge invariance of the Hamiltonian imposes the following constraint on the system,
\beq
\frac{1}{4\pi}{\bm \nabla}\cdot {E}(x) = -i2e\left[
\pi(x)\Delta^{\ast}(x) - \pi^{\ast}(x) \Delta(x)
\right].
\eeq
This is the Hamiltonian representation of Gauss's law. If the initial conditions satisfy Gauss's law, this law will hold at later times. As explained below, for numerical simulations, we discretize the Hamiltonian. The discretized Hamiltonian maintains the discrete version of the ${\rm U}(1)$ gauge transformations, ensuring that the equations of motion exactly conserve electric charge throughout. In the Lagrangian formalism, however, a numerical scheme entails finite differencing the Euler-Lagrange equations and violates Gauss's law and charge conservation.  

All computations are carried out on grids that are either two-dimensional or three-dimensional. The mesh width along the $\mu$-direction ($\mu=x,y,z$) is set to $a_{\mu}$, and the neighboring lattice points are connected by links. The fields $\Delta$ is evaluated on the grids as $\Delta_{i,j,k}(t) \equiv \Delta({\bm x}_{\bm i},t)$, while the vector potential ${A}_{\mu}$ is defined on the link conneting ${\bm x}_{\bm i}$ and ${\bm x}_{\bm i}+\hat{\bm e}_{\mu}a_{\mu}$ as $A^{i,j,k}_{\mu}(t) \equiv A_{\mu}({\bm x}_{\bm i},t)$. Here, ${\bm x}_{\bm i}\equiv (x_i,y_j,z_k)$ denotes the position of the lattice site ${\bm i}=(i,j,k)$ and $\hat{\bm e}_{\mu}$ is a unit vector along the $\mu$-drection, where $i=1,\cdots,N_x$, $j=1,\cdots,N_y$, and $k=1,\cdots,N_z$. We introduce the link variables as
\begin{align}
U^{\bm i}_{\mu}[A_{\mu}]\equiv& \exp\left(
-i\frac{2e}{c}\int^{{\bm x}_{\bm i}+a_{\mu}\hat{\bm e}_{\mu}}_{{\bm x}_{\bm i}} {\bm A}({\bm x},t)\cdot d{\bm x}
\right) \nn \\
\approx& e^{
-i\frac{2e}{c}a_{\mu} A^{\bm i}_{\mu}(t)}, 
\end{align}
and $\bar{U}^{\bm i}_{\mu}[A_{\mu}] = (U^{\bm i}_{\mu}[A_{\mu}])^{\ast}$. The superconducting order parameter ($\Delta$ and $\pi$) and gauge fields (${\bm A}$ and ${\bm \varepsilon}$) are defined on each lattice site and a link connecting neighboring sites, respectively. We also introduce the gauge-invariant variable as
\begin{align}
W_z^{i,j,k}[{\bm A}]
&\equiv U_x^{i,j,k}(t)U_y^{i+1,j,k}(t)\bar{U}_{x}^{i,j+1,k}(t)\bar{U}_y^{i,j,k}(t) \nn \\
&= e^{-i\frac{2e}{c}a_xa_y B^{i,j,k}_z(t)}.
\end{align}
The product of four link variables is related to the magnetic flux penetrating the plaquette. By cyclic permutation, $(x,y,z)\rightarrow (y,z,x)$ and $(i,j,k)\rightarrow(j,k,i)$, the expressions for $W_y^{\bm i}(t)$ and $W_z^{\bm i}(t)$ can be derived.

In the temporal gauge, the dynamics of $\Delta_{\bm i}(t)$, ${\bm A}^{\bm i}_{\mu}(t)$, $\pi_{\bm i}(t)$, and $\varepsilon^{\bm i}_{\mu}(t)$ are governed by the Hamilton's equations, $\dot{\Delta}_{\bm i}=\partial\mathcal{H}/\partial \pi_{\bm i}$, $\dot{A}^{\bm i}_{\mu}=\partial\mathcal{H}/\partial{\varepsilon}^{\bm i}_{\mu}$, $\dot{\pi}_{\bm i}=-\partial\mathcal{H}/\partial \Delta_{\bm i}$, and $\dot{\varepsilon}^{\bm i}_{\mu}=-\dot{E}^{\bm i}_{\mu}/4\pi c=-\partial\mathcal{H}/\partial A^{\bm i}_{\mu}$, which read
\begin{align}
\frac{d\Delta_{\bm i}}{dt} = -\frac{1}{\tau}{\pi}_{\bm i}, \quad
\frac{dU_{\mu}^{\bm i}}{d t} = -i U^{\bm i}_{\mu}[A_{\mu}^{\bm i}]\tilde{E}^{\bm i}_{\mu}, 
\label{eq:Delta_U}
\end{align}
and
\begin{align}
\frac{d\pi_{\bm i}}{dt} 
=& -\frac{\kappa}{a_{\mu}^2}
\sum_{\mu}\bigg\{U^{{\bm i}+\hat{\bm e}_{\mu}}_{\mu}[\mathcal{A}_{\mu}^{{\bm i}+\hat{\bm e}_{\mu}}]
\Delta_{{\bm i}+\hat{\bm e}_{\mu}}
+\bar{U}^{{\bm i}-\hat{\bm e}_{\mu}}_{\mu}[\mathcal{A}_{\mu}^{{\bm i}+\hat{\bm e}_{\mu}}]
\Delta_{{\bm i}-\hat{\bm e}_{\mu}}\nn \\
&-2\Delta_{\bm i} \bigg\}
+ \beta\left(|\Delta_{\bm i}|^2-\Delta^2_0\right)\Delta_{\bm i}, 
\label{eq:pi} \\
\frac{d\tilde{E}^{\bm i}_{x}}{dt} = &  
-c^2{\rm Im}\bigg[
\frac{U^{\bm i}_x[\mathcal{A}_{x}^{\bm i}]\Delta^{\ast}_{\bm i}\Delta_{{\bm i}+\hat{\bm e}_x}}{\lambda^2\Delta^2_0} 
+\frac{W^{\bm i}_z[{\bm A}^{\bm i}]-W^{{\bm i}-\hat{\bm e}_y}_z[{\bm A}^{{\bm i}-\hat{\bm e}_y}]}{a_y^2} \nn \\
&-\frac{W^{\bm i}_y[{\bm A}^{\bm i}]-W^{{\bm i}-\hat{\bm e}_z}_y[{\bm A}^{{\bm i}-\hat{\bm e}_z}]}{a_z^2}
\bigg],
\label{eq:epsilonx}
\end{align}
where $\lambda^{-2}\equiv 32e^2\pi\Delta^2_0/c^2$ is the magnetic penetration length, and we have introduced $\tilde{E}^{\bm i}_{\mu}\equiv 2e a_{\mu}E^{\bm i}_{\mu}$. 
The equations of motion for ${E}^{\bm i}_{y}$ and  ${E}^{\bm i}_{z}$ are obtained from Eq.~\eqref{eq:epsilonx} by cyclic permutation of subscripts.

\subsection{S1.3 Boundary conditions in three-dimensional caclulations}

\begin{figure}[t!]
\includegraphics[width=85mm]{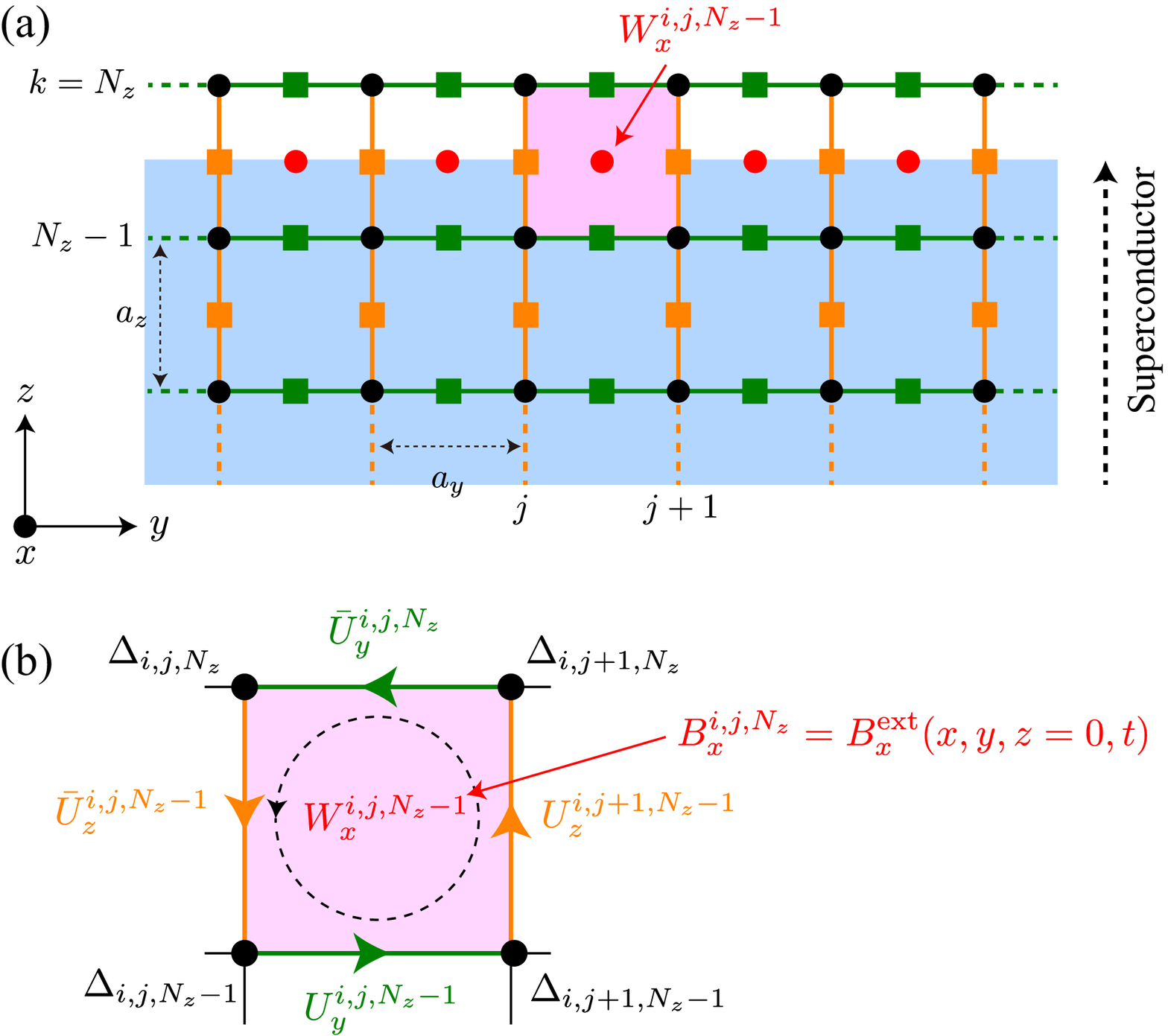}
\caption{(a) Computation grids for $\Delta_{\bm i}$, $\pi _{\bm i}$ (filled circles), $U^{\bm i}_{\mu}$, $\tilde{\varepsilon}_{\mu}^{\bm i}$ (squares). (b) The link variable $U_y^{i,j,N_z}$ is determined by the magnetic flux generated by the vortex beam through the plaquette, ${B}^{\rm ext}_xa_ya_z$.}
\label{fig:grid}
\end{figure}

Here we define the boundary conditions for calculating Eqs.~\eqref{eq:Delta_U}-\eqref{eq:epsilonx}. Any current passing through the boundary between a superconductor and a vacuum is zero. Thus we impose ${\bm j}\cdot\hat{\bm n}=0$ on the boundary of the superconductor, $\partial\Omega$, where $\hat{\bm n}$ is a unit vector normal to $\partial \Omega$. This requires the boundary condition for the condensate as $\Delta_{1,j,k}=U_x^{1,j,k}\Delta_{2,j,k}$ and $\Delta_{N_x,j,k}=\bar{U}_x^{N_x-1,j,k}\Delta_{N_x-1,j,k}$, and a similar condition applies for the $y$ and $z$ directions.

For numerical simulations in two dimensions, the effect of the irradiated vortex beam is taken into account as the external field ${\bm A}^{\rm ext}(x)$ through Eq.~\eqref{eq:Atotal}. 
In the case of three-dimensional superconducting films, the effect is incorporated through the boundary condition on the upper surface of the three-dimensional superconductor. The values of ${\bm A}$ on the upper surface ($k=N_z$) are determined in a way that satisfy ${\bm \nabla}\times {\bm A} = {\bm B}^{\rm ext}$. This ensures that the value of the internal magnetic field matches a magnetic field generated by an vortex beam ${\bm B}^{\rm ext}$. Figure~\ref{fig:grid} illustrates how to implement the boundary conditions on the discretized grids. For instance, the boundary value of the link variable $U_y^{i,j,N_z}$ on the upper surface is obtained from $B_x^{\rm ext}$ as
\beq
U_y^{i,j,N_z} = U_y^{i,j,N_z-1}U_z^{i,j+1,N_z-1}\bar{U}_{z}^{i,j,N_z-1}e^{i\frac{2e}{c}a_xB_x^{\rm ext}}.
\eeq
The cyclic permutation of subscripts gives the boundary conditions on $U_z^{i,j,N_z}$ and $U_x^{i,j,N_z}$. Hence, the EM fields generated by vortex beams can be implemented as the boundary condition for three-dimensional numerical simulations.

\subsection{S1.4 Laguerre-Gaussian vortex beams}

In this work, we examine the effects of irradiating vortex beams, which are a type of topological light that possess both orbital and spin angular momentum (OAM and SAM, respectively), on $s$-wave superconductors. In the Lorenz gauge, the vector potential obeys the Helmholtz equation
\beq
\left(
{\bm \nabla^2}+k^2
\right){\bm A}^{\rm L}({\bm x})={\bm 0},
\eeq
where we assume the monochromatic wave, ${\bm A}^{\rm L}({\bm x},t) ={\bm A}^{\rm L}({\bm x}) e^{-i\Omega t}$ with $\Omega = ck$. The superscript ``L'' indicates the Lorenz gauge.
Let us examine an EM field that propagates along the $z$-axis. This field is represented by 
${\bm A}^{\rm L}({\bm x}) = u({\bm x})e^{ikz}\hat{\bm e}_{s}$, where $\hat{\bm e}_{s}= (\hat{\bm e}_x+is\hat{\bm e}_y)/\sqrt{2}$ is the polarization vector of the circularly polarized light and the helicity $s = \pm 1$ corresponds to the SAM of light. The scalar potential $\phi^{\rm L}$ is related to the vector potential ${\bm A}^{\rm L}$ through the Lorenz condition, ${\bm \nabla}\cdot{\bm A}^{\rm L}-\frac{ik}{c}\phi^{\rm L}=0$. We define the wave vector in the cylindrical coordinate as ${\bm k}= k_{\perp}\cos\theta_{k}\hat{\bm e}_x + k_{\perp}\sin\theta_{k}\hat{\bm e}_x + k_z\hat{\bm e}_z$. To solve the Helmholtz equation, we use the paraxial approximation, which satisfies the conditions, $| \frac{\partial^2u}{\partial z^2}|\ll | \frac{\partial^2u}{\partial x^2}|$, $|\frac{\partial^2u}{\partial y^2}|$, and $| \frac{\partial^2u}{\partial z^2}|\ll 2 k| \frac{\partial u}{\partial z}|$. This indicates that the amplitude distribution of the vector potential changes slowly with distance $z$ when compared to variations of $u({\bm x})$ in the lateral direction. Then, 
the second-order derivative term can be removed from the Helmholtz equation, 
\beq
\left( {\bm \nabla}^2_{\perp} + 2ik\frac{\partial}{\partial z} \right)u({\bm x})=0.
\label{eq:Helm}
\eeq
In the paraxial approximation, the cylindrically symmetric solution describes the Laguerre-Gaussian mode
\begin{align}
u_{p,m}({\bm x}) =& \frac{1}{w(z)}\left(\frac{\sqrt{2}\rho}{w(z)}\right)^{|m|}e^{-\rho^2/w^2(z)}L^{|m|}_p\left( 
\frac{2\rho^2}{w^2_0}
\right)e^{im\theta}\nn \\
&\times \exp\left(\frac{ik\rho^2z}{2(z^2+z^2_{\rm R})}-i(2p+m+1)\tan^{-1}\!\frac{z}{z_{\rm R}}\right),
\label{eq:uLG}
\end{align}
where $L^{m}_p$ is the associated Laguerre polynomial and ${\bm x}=(\rho,\theta,z)$ is the cylindrical coordinate. The beam waist $w_0$ determines the width, $w(z)=w_0\sqrt{1+z^2/z^2_{\rm R}}$, and the Rayleigh range of the beam, $z_{\rm R} = kw^2_0/2$. At the focal plane ($z=0$), the LG mode reduces to 
\begin{align}
u_{p,m}(\rho,\theta,z=0) =& \frac{1}{w_0}\left(\frac{\sqrt{2}\rho}{w_0}\right)^{|m|}e^{-\rho^2/w^2_0}L^{|m|}_p\left( 
\frac{2\rho^2}{w^2_0}
\right)e^{im\theta}.
\end{align}

\begin{figure}[t!]
\includegraphics[width=85mm]{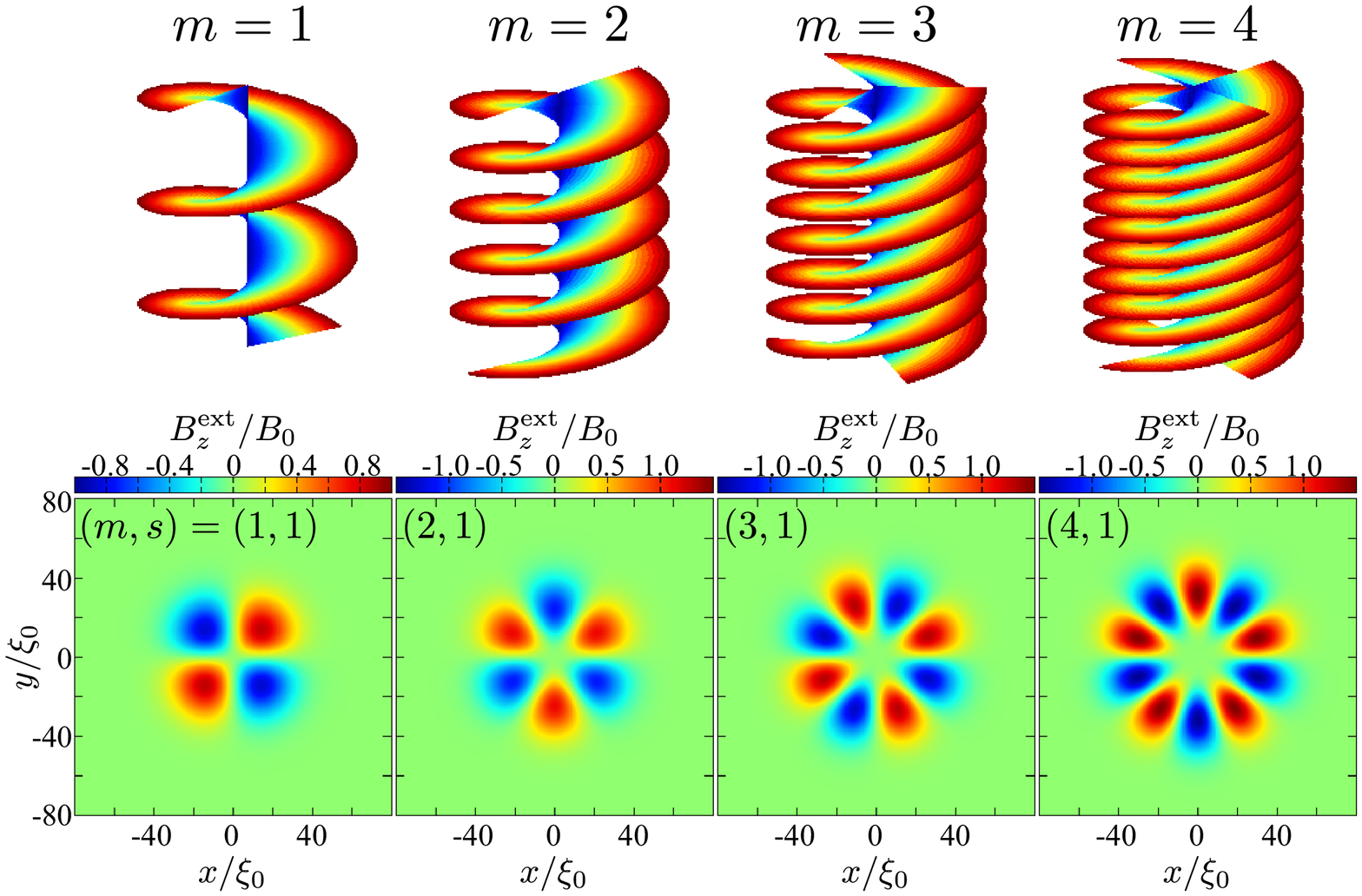}
\caption{Upper panels show the wavefronts of vortex beams with $m=1$, $2$, $3$, and $4$ (from left to right). Lower panels show the snapshots of the spatial profile of the longitudinal magnetic field ($B^{\rm ext}_z$) at $t=0$ generated by the corresponding vortex beams. The oscillation pattern reflects the total angular momentum of light, $J=m+s$.}
\label{fig:beams}
\end{figure}

The Laguerre-Gaussian mode has two integers, $p$ and $m$. The phase twist $e^{im\theta}$ is the eigenvalue of the OAM operator $L_z=-i\partial_{\theta}$. The topological charge, $m$, counts the number of phase windings in a single wavelength. The wavefront, or the isophase plane, takes a spiral shape around the propagation axis, as depicted in Fig.~\ref{fig:beams}. To ensure the single-valuedness of the EM fields, $u_{p,m}({\bm x})$ must equal zero at the center of the beam ($\rho=0$). This results in a doughnut-shaped intensity profile of the vortex beams in the transverse plane. Similar to a quantum vortex in superfluids, the vortex beam carries OAM ($m$) in addition to SAM ($s$). 

Using the Laguerre-Gaussian mode in Eq.~\eqref{eq:uLG} and the Lorenz condition, one derives the electric field ${\bm E}^{\rm ext}_{p,m}=i\Omega {\bm A}^{\rm L}-{\nabla}\phi^{\rm L}$ in relation to ${\bm A}^{\rm L}$ as 
\beq
{\bm E}^{\rm ext}_{p,m}({\bm x}) = i\Omega\left[
\hat{\bm e}_{s}u_{p,m}({\bm x})+ \hat{\bm z}\frac{i}{k}\hat{\bm e}_s \cdot{\bm \nabla}_{\perp}u_{p,m}({\bm x})
\right]e^{ikz},
\label{eq:ELG}
\eeq
where ${\bm \nabla}_{\perp}\equiv (\partial_x,\partial_y,0)$. In the same manner, the magnetic field ${\bm B}^{\rm ext}={\bm \nabla}\times {\bm A}^{\rm L}$ reads 
\beq
{\bm B}^{\rm ext}_{p,m}({\bm x}) = ik\left[
\hat{\bm z}\times\hat{\bm e}_{s}u_{p,m}({\bm x})
+ \frac{i}{k}\hat{\bm e}_s\times {\bm \nabla}_{\perp}u_{p,m}({\bm x})
\right].
\label{eq:bext}
\eeq
To obtain the vector potential in the temporal gauge, ${\bm A}^{\rm ext}$, we compare Eq.~\eqref{eq:ELG} with the equation for the electric field in the temporal gauge, ${\bm E}^{\rm ext}=i\Omega {\bm A}^{\rm ext}$. Then, the vector potential in the temporal gauge reads
\beq
{\bm A}^{\rm ext}({\bm x})=\left(\hat{\bm e}_{s}u_{p,m}({\bm x})+ \hat{\bm z}\frac{i}{k}\hat{\bm e}_s \cdot{\bm \nabla}_{\perp}u_{p,m}({\bm x})\right)e^{ikz}.
\label{eq:Aext}
\eeq
In Fig.~\ref{fig:beams}, the bottom panels display the spatial profile of the longitudinal magnetic field ($B_z$) generated by vortex beams with $(m,s)=(1,1)$, $(2,1)$, $(3,1)$, and $(4,1)$ and $p=0$. The oscillation pattern on the circumference indicates the total angular momentum of the beams, which is denoted by $J=m+s$.

We now represent the vector potential of the Laguerre-Gaussian beam as a superposition of plane waves 
\beq
{\bm A}^{\rm ext}_{p,m}({\bm x}) = \int \frac{d{\bm k}_{\perp}}{(2\pi)^2}e^{i{\bm k}_{\perp}\cdot{\bm r}_{\perp}}{\bm A}^{\rm ext}_{p,m}({\bm k}_{\perp}).
\eeq
The Fourier coefficient of the amplitude distribution at the focal plane is given by
\beq
{\bm A}^{\rm ext}_{p,m}({\bm k}_{\perp}) 
= \int^{\infty}_0\rho d\rho \int^{2\pi}_0 d\theta {\bm A}^{\rm ext}_{p,m}(\rho,\theta)e^{-ik_{\perp}\rho\cos(\theta-\theta_k)}.
\label{eq:U}
\eeq
To calculate the integral over the angle $\theta$, we use the integral representation of the Bessel function of the first kind, $J_m(x)$, 
\beq
J_m(x) = (\mp i)^m e^{-im\theta_k}\int^{2\pi}_0 \frac{d\theta}{2\pi} e^{im\phi\pm ix\cos(\theta-\theta_k)}.
\label{eq:JJ}
\eeq
To this end, the Fourier component of the Laguerre-Gaussian beam is given by~\cite{pes17}
\begin{align}
{\bm A}^{\rm ext}_{p,m}({\bm k}_{\perp}) 
&= \hat{\bm e}_{s}(-i)^m\pi w_0 e^{im\theta_k}e^{-k^2_{\perp}w^2_0/4} \left(
\frac{k_{\perp}w_0}{2}
\right)^{|m|}\nn \\
& \times \sum^p_{\beta=0}(-1)^{\beta}2^{\beta+|m|/2} 
\begin{pmatrix}
p+m \\ p-\beta
\end{pmatrix}
L^{|m|}_{\beta}\left(
\frac{k^2_{\perp}w^2_0}{4}
\right).
\label{eq:ap}
\end{align}
When the OAM is not zero ($m\neq 0$), vortex beams exhibit a distinctive intensity pattern in momentum space that resembles a doughnut shape. The intensity peaks around $k_{\perp}\sim 2/w_0$.


\section{S2. Numerical simulations in two dimensions}

\begin{figure*}[t!]
\includegraphics[width=180mm]{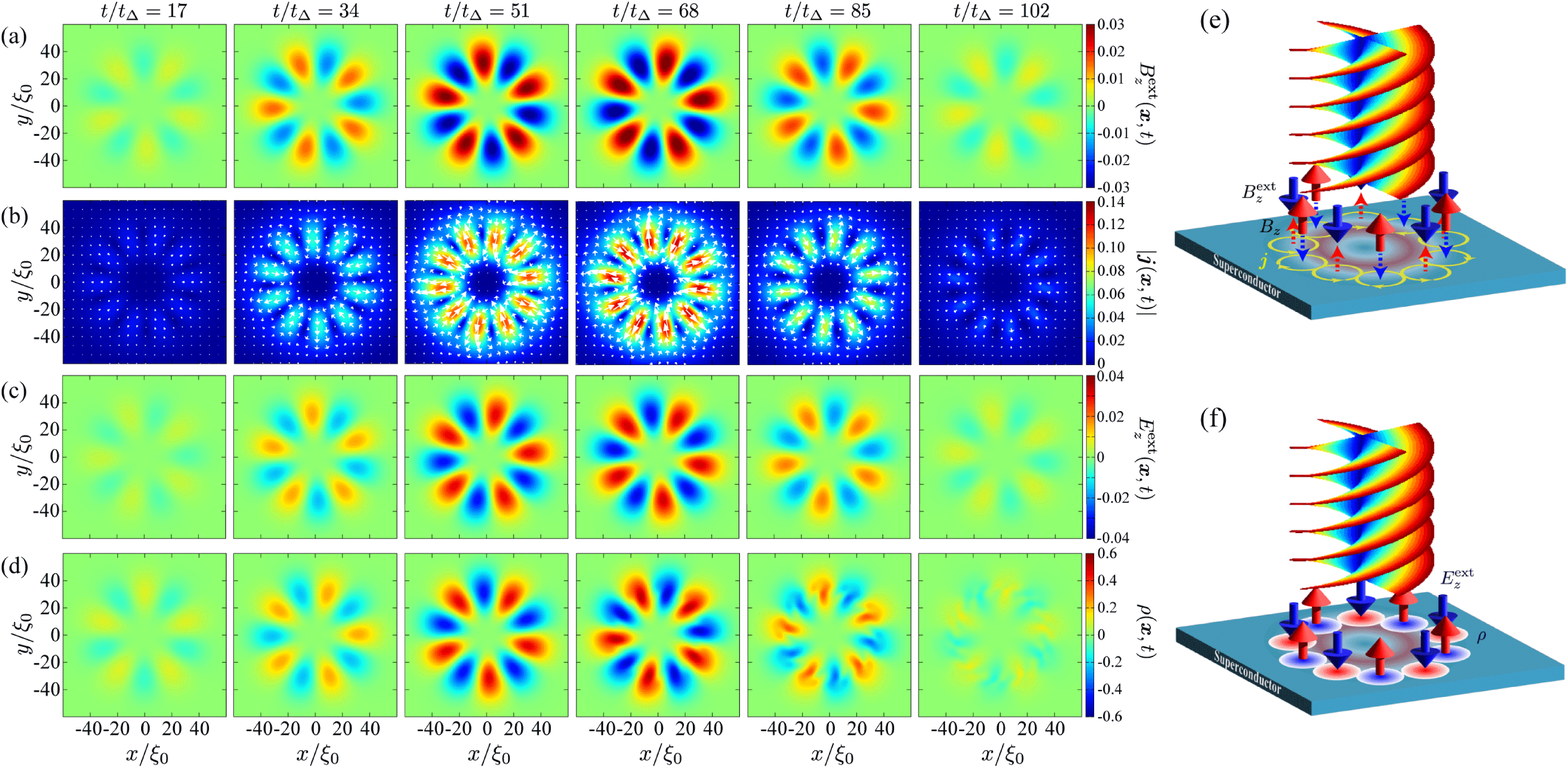}
\caption{Snapshots of $B^{\rm ext}_z({\bm x},t)$ (a), $|{\bm j}({\bm x},t)|$ (b), $E^{\rm ext}_z({\bm x},t)$ (c) and $\rho({\bm x},t)$ (d) at $t/t_{\Delta} = 17$, 34, 51, 68, 85, and 102 for the pulsed vortex beam with $\Omega = 1.025\Delta_0$ and $(m,s)=(4,1)$. Here $E^{\rm ext}_z$ and $B^{\rm ext}_z$ are the longitudinal electric and magnetic fields of the vortex beam, respectively. In (b), the arrows denote the vector field of the supercurrent density ${\bm j}({\bm x},t)$. (e,f) Schematics that the longitudinal magnetic and electric fields ($B^{\rm ext}_z$ and $E^{\rm ext}_z$) by the vortex beam with $m\neq 0$ involve the eddy supercurrent density (${\bm j}$) and charge density oscillation ($\rho$), respectively. In (e), the eddy supercurrent induces the alternative magnetic fields which screens $B^{\rm ext}_z$.
}
\label{fig:snap2d}
\end{figure*}

Here we consider the dynamics of the superconducting order and EM fields after the irradiation of a pulsed vortex beam.
The vector potential for the pulsed vortex beam with $(m,s)$ is given by
\beq
{\bm A}^{\rm ext}({\bm x},t) = {\rm Re}\left\{\frac{A_0u_{p,m}({\bm x})}{\max | u_{p,m}({\bm x})|}
\exp\left[
-\left( \frac{t-t_{0}}{\sigma}\right)^2 -i\Omega t
\right]\hat{\bm e}_s\right\},
\label{eq:pulse}
\eeq
where $\Omega$ and $\sigma\equiv 2\pi n_{\rm p}/\Omega$ are the frequency and the full width at half-maximum of the beam intensity, respectively. The pulsed field reaches the maximum intensity at $t=t_0=2\sigma$. In the main text, we focus on a set of parameters: The number of cycles of the pulse field $n_{\rm p}=5$, the Ginzburg-Landau (GL) parameter $\kappa_{\rm GL}\equiv \lambda /\xi_0 =10$, and the beam waist $w_0=20\xi_0$. 
In both the main text and the supplemental materials, we mainly show the numerical results for $\Omega = 1.025\Delta_0$ as it is the frequency at which the nonlinear resonance of light to the Higgs excitation occurs. While the conventional Gaussian beam resonates at $2\Omega = 2\Delta_0$, the resonance frequency shifts from $2\Delta_0$ due to the spatial inhomogeneity of the vortex beam.
In this section, we discuss how these parameters have an impact on the nonlinear dynamics of the superconducting order and gauge fields caused by vortex beams. The remaining parameters are identical to those mentioned in the main text. 

In this section, we present the numerical simulations and analysis with the temporal gauge. However, the main results remain consistent regardless of the choice of the gauge. In fact, the same logical flow shown below is obtained from the Coulomb gauge.

\subsection{S2.1 Roles of longitudinal magnetic and electric fields}

Vortex beams with $m\neq 0$ involve both the longitudinal magnetic and electric fields. As mentioned in the main text, the OAM of light, $m$, increases the THG intensity. This is caused by the phase and charge excitations that transfer the OAM of light to the condensate and amplify the nonlinear coupling of the Higgs mode with light. It is important to note that the longitudinal fields play a crucial role in facilitating the charge and phase excitations.

In Fig.~\ref{fig:snap2d}(a) and \ref{fig:snap2d}(b), we show the snapshots of the longitudinal magnetic field of the vortex beam $B^{\rm ext}_z({\bm x},t)$ and the current density ${\bm j}({\bm x},t)$, respectively. The longitudinal magnetic field oscillates in the azimuthal direction with the total angular momentum of light, $m+s=5$. The current density in Fig.~\ref{fig:snap2d}(b) features a periodic array of eddy currents in the azimuthal direction, which generates the staggered internal field. Figure \ref{fig:snap2d}(e) shows the schematic picture that the longitudinal magnetic field is screened by the staggered internal field generated by the eddy supercurrent. The spatially inhomogeneous current density also causes the oscillation of the charge density as $\partial_t \rho ({\bm x},t) = -{\bm \nabla}\cdot{\bm j}({\bm x},t)\neq 0$. As the charge density is related to the oscillation of the phase mode, $\rho \propto \partial _t \varphi$, in our choice of the gauge, the longitudinal magnetic field of the vortex beam can cause the phase mode excitation. The oscillation of the charge density also follows the longitudinal electric field generated by the vortex beam as seen in Figs.~\ref{fig:snap2d}(c), \ref{fig:snap2d}(d), and \ref{fig:snap2d}(f). Hence, the longitudinal magnetic field generated by vortex beams with $m\neq 0$ induces both a periodic array of eddy currents and the charge oscillation through the continuity equation. Instead of the longitudinal magnetic field, this is also understandable with the longitudinal {\it electric} field displayed in Fig.~\ref{fig:snap2d}(c). The electric field excites the charge density oscillation and characteristic current distribution. Conventional Gaussian beams, which correspond to light with $m=0$ and have no longitudinal component, cannot cause phase oscillations.

\begin{figure}[t!]
\includegraphics[width=85mm]{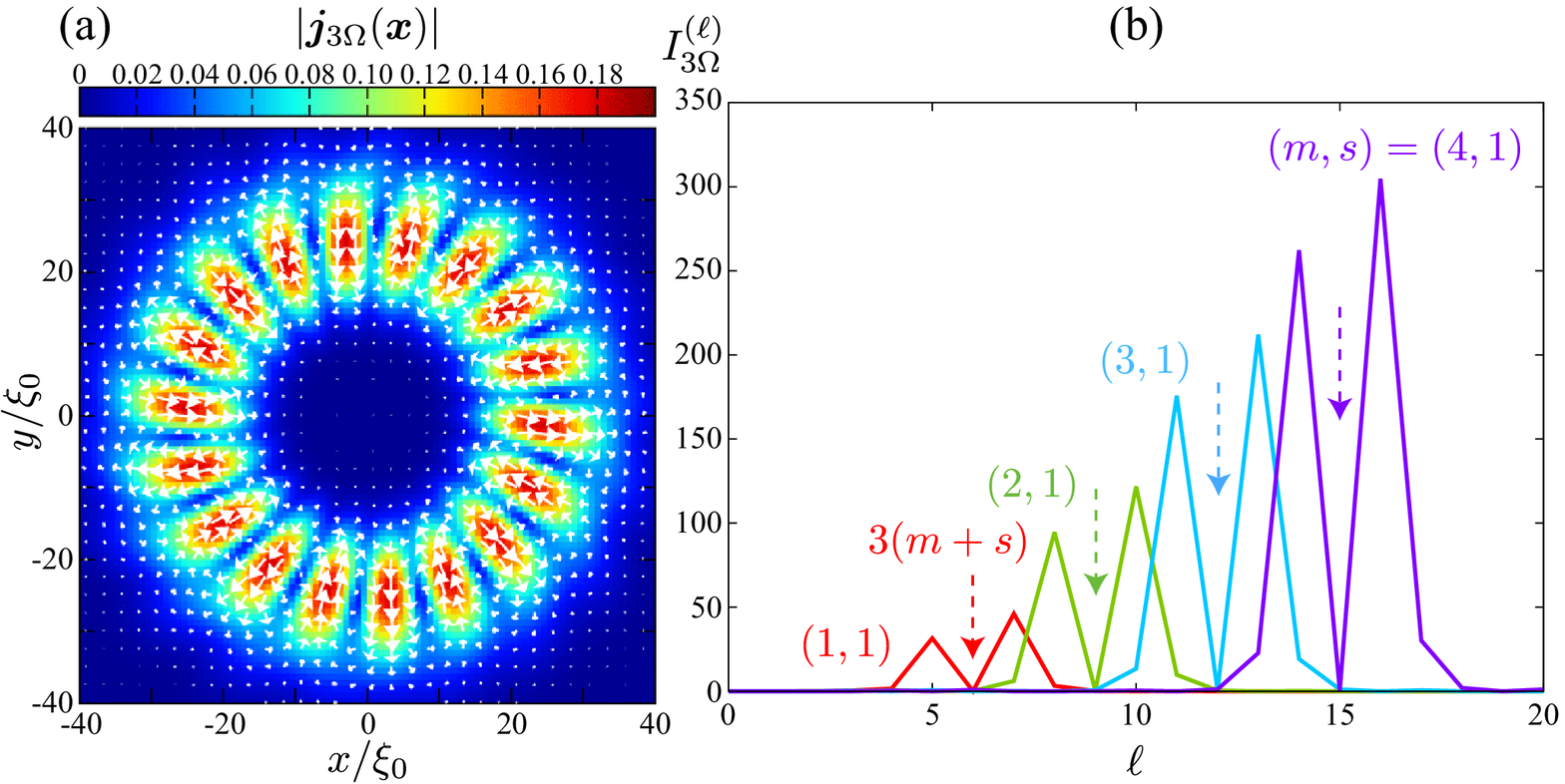}
\caption{(a) Spatial profile of the $3\Omega$ component of the supercurrent density, ${\bm j}_{3\Omega}({\bm x})\equiv \int dt {\bm j}({\bm x},t)e^{i3\Omega t}$, for the vortex beam with $\Omega = 1.025\Delta_0$ and $(m,s)=(2,1)$, where the color map and vector field represent $|{\bm j}_{3\Omega}({\bm x})|$ and ${\bm j}_{3\Omega}({\bm x})$, respectively. (b) Projection of the third-order current response to the angular momentum eigenstates, $I_{3\Omega}^{(\ell)}$ for $m=1$, 2, 3, 4. Each arrow shows three times the total angular momentum of light, $3(m+s)$.
}
\label{fig:fourier}
\end{figure}

\subsection{S2.2 Spatial profiles of third-order current response}

Here we clarify the spatial profiles of the third-order current response to vortex beams, which is defined by ${\bm j}_{3\Omega}({\bm x}) \equiv \int dt {\bm j}({\bm x},t)e^{i3\Omega t}$. This nonlinear response ${\bm j}_{3\Omega}$ is associated with the THG intensity. In Fig.~\ref{fig:fourier}(a), we plot the spatial profile of ${\bm j}_{3\Omega}({\bm x})$ for the vortex beam with $(m,s)=(2,1)$. The current density consists of a periodic array of the eddy current along the azimuthal direction, and its oscillation pattern reflects three times the total angular momentum of light, $3(m+s)$. We expand the third-order current response in terms of the eigenstates of the angular momentum and the coefficient, $I_{3\Omega}^{(\ell)}\equiv \int d{\bm x} e^{-i\ell \theta}{\bm j}_{3\Omega}({\bm x},t)$ ($\ell\in \mathbb{Z}$), is plotted in Fig~\ref{fig:fourier}(b). This peaks around three times the total angular momentum of light, $3(m+s)$.

\begin{figure}[t!]
\includegraphics[width=85mm]{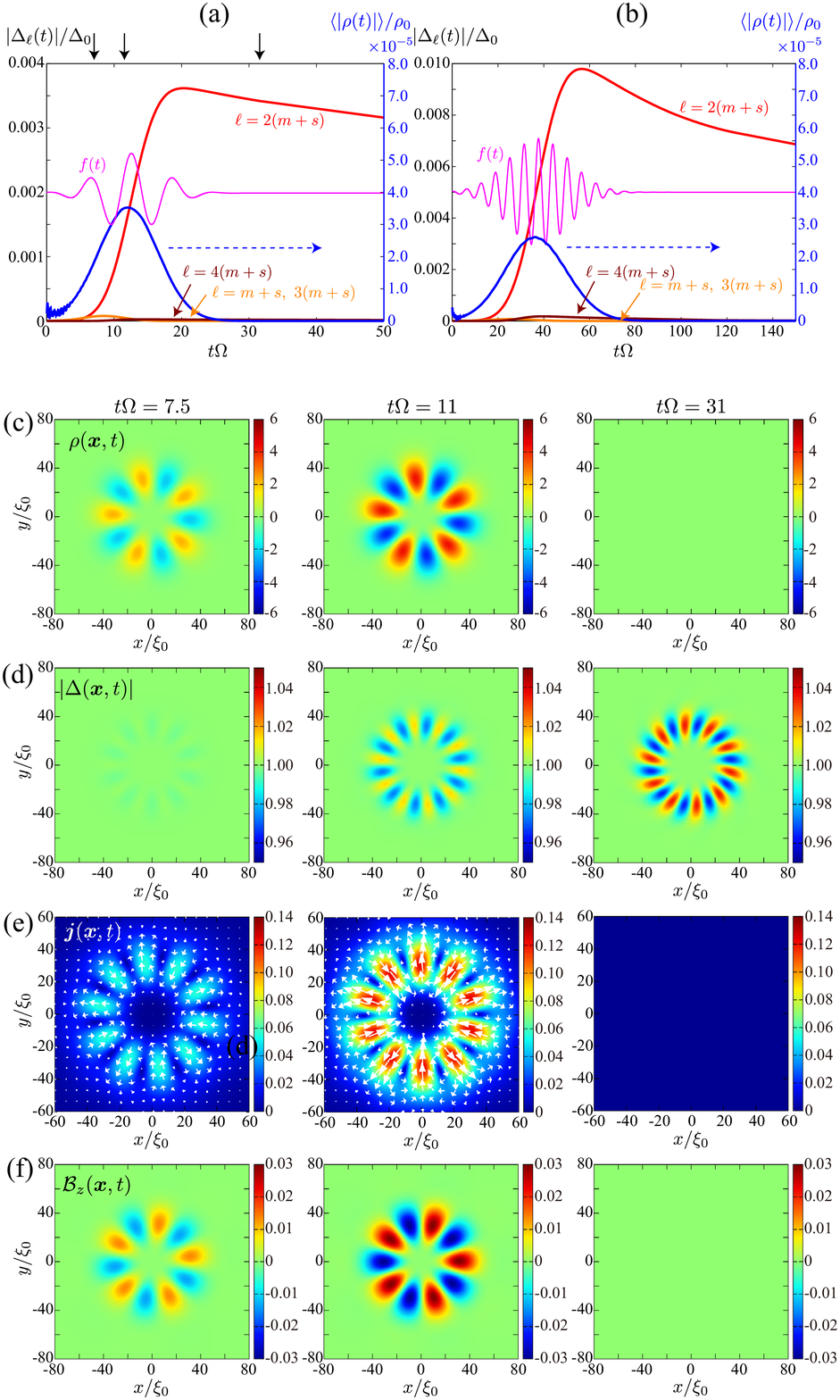}
\caption{(a,b) Time evolutions of $\Delta_{\ell=2J}$ for the single-cycle pulse $n_{\rm p}=1$ (a) and $n_{\rm p}=3$ (b), where the magenta curve shows the temporal profile of the applied field $f(t)=\cos(\Omega t)e^{-(t-t_0)^2/\sigma^2_t}$. The averaged charge density $\langle |\rho(t)|\rangle$ is also plotted in (a) and (b). In (a), the arrows (black) point to the time, $t\Omega=7.5$, $11$, and $31$, corresponding to the snapshots displayed in (c-f). (c-f) Snapshots of $\rho({\bm x},t)$ (c), $|\Delta({\bm x},t)|$ (d), ${\bm j}({\bm x},t)$ (e), and $\mathcal{B}_z({\bm x},t)$ (f) at $t\Omega=7.5$, $11$, and $31$ for $n_{\rm p}=1$, where $\mathcal{B}_z$ is the sum of the internal field $B_z\equiv {\bm \nabla}\times {\bm A}$ and the external one ${\bm B}^{\rm ext} = {\bm \nabla}\times {\bm A}^{\rm ext}$. In (e), the vectorial field corresponds to $(j_x,j_y)$, and the color map is $|{\bm j}|$. For all data, we set $(m,s)=(4,1)$, $\Omega/\Delta_0=1.025$, and $\kappa_{\rm GL}=10$.}
\label{fig:pulse}
\end{figure}

\subsection{S2.3 Number of cycles of the pulse field} 

We demonstrate that the fundamental characteristics of the dynamics of superconductors induced by vortex beams, including the transient charge density oscillation and spiral Higgs waves, are not influenced by the number of cycles. Figures~\ref{fig:pulse}(a) and \ref{fig:pulse}(b) show the time evolution of $\Delta_{\ell=2J}$ and the averaged charge density $\langle |\rho(t)|\rangle\equiv \int d{\bm x}|\rho({\bm x},t)|$ for $n_{\rm p}=1$ and $n_{\rm p}=3$, respectively. The OAM and SAM of light are set to $(m,s)=(4,1)$. We expand the superconducting order parameter in terms of the eigenstates of the angular momentum, $\Delta_{\ell}\equiv \int d{\bm x} e^{i\ell \theta}|\Delta({\bm x},t)| $ with $\ell\in\mathbb{Z}$. The vortex beam with the total angular momentum $J=m+s$ induces the nonlinear excitation of $\ell = 2J$ component via the coupling of the order parameter amplitude $|\Delta({\bm x},t)|$ to the gauge-invariant potential $\tilde{\bm{\mathcal{A}}}^2$. As shown in Fig.~2(b) of the main text, which displays the results for $n_{\rm p}=5$, the pulsed vortex beam transiently excites the charge density. The charge density returns to neutral distribution after the pulse is turned off. In Fig.~\ref{fig:pulse}(c), we present the snapshots of $\rho({\bm x},t)$ at $t\Omega = 7.5$, $11$, and $31$ for $n_{\rm p}=1$. The oscillation pattern along the azimuthal direction reflects the total angular momentum $J=m+s=5$. As shown in Fig.~\ref{fig:pulse}(a), the $\ell=2J$ component of the condensate, $\Delta_{\ell=2J}$, sharply increases around $t\sim t_0 = 4\pi\Omega^{-1}$ and persists even after the pulse ends. Figure~\ref{fig:pulse}(d) shows the standing wave on the condensate, reflecting twice the total angular momentum, $|\Delta({\bm x},t)|\propto \sin(2J\theta-\Omega t)$. We also dislay snapshots of the current density ${\bm j}({\bm x},t)$ and the total magnetic field $\mathcal{B}_z({\bm x},t)$ in Figs.~\ref{fig:pulse}(e) and \ref{fig:pulse}(f), respectively. The current response consists of a periodic array of the clockwise and counter-clockwise eddy currents that generate the internal magnetic field to screen $B_z^{\rm ext}$.

\subsection{S2.4 Screening effect and $\kappa_{\rm GL}$ dependence} 

Let us now examine the screening effect and the $\kappa_{\rm GL}$ dependence of the nonlinear Higgs response. We focus on a two-dimensional superconducting film that is thinner than the London penetration depth. In this situation, the transverse components of the external fields $({A}^{\rm ext}_x,{A}^{\rm ext}_y)$ couple to the superconducting film through the minimal coupling terms. The Maxwell-Amp\`{e}re law in Eq.~\eqref{eq:ampere} generates the internal fields $({A}_x,{A}_y)$. The transverse components $({A}^{\rm ext}_x,{A}^{\rm ext}_y)$ are screened by the internal field, while the longitudinal component $A^{\rm ext}_z$ remains unscreened.

To understand the screening effect, let us examine the spatial profiles of the current response [Fig.~\ref{fig:pulse}(e)] and the current-induced magnetic field profiles [Fig.~\ref{fig:pulse}(f)]. These characteristic profiles can be explained through the linear response analysis of the coupled equations for $\varphi(x)$ and $\bm{\mathcal{A}}(x)$. The continuity equation \eqref{eq:cont} gives the equation of motion for $\varphi(x)$, which can be simplified to
\beq
\left( {\bm \nabla}^2_{\perp} - \frac{1}{v^2}\frac{\partial^2}{\partial t^2} \right)\varphi(x) = \frac{2e}{c}{\bm \nabla}_{\perp}\cdot\bm{\mathcal{A}}(x),
\label{eq:eom_phase}
\eeq
in the linear order to ${\bm A}^{\rm ext}$, where ${\bm \nabla}_{\perp}\equiv(\partial_x,\partial_y,0)$ is the derivatives in two dimension. The right-hand side of Eq.~\eqref{eq:eom_phase} is attributed to the interaction of light with the two-dimensional superconducting thin film, resulting in ${\bm \nabla}_{\perp}\cdot{\bm A}^{\rm ext}$, rather than ${\bm \nabla}\cdot{\bm A}^{\rm ext}$, where the latter goes to zero. As shown in Eq.~\eqref{eq:Aext}, ${\bm \nabla}_{\perp}\cdot{\bm A}^{\rm ext}\neq 0$ is a unique character of vortex beams with nonzero OAM and $A^{\rm ext}_z\neq 0$ in contrast to conventional Gaussian beams. This is essential for the off-resonant excitation of the phase mode.

In the temporal gauge, the dynamics of ${\bm A}$ is governed by Eqs.~\eqref{eq:ampere} and \eqref{eq:current}. The equation reduces to 
\begin{align}
{\bm \nabla}_{\perp}\times {\bm \nabla}_{\perp}\times {\bm A}(x)+\frac{1}{c^2}\frac{\partial^2{\bm A}(x)}{\partial t^2} = -\frac{1}{\lambda^2}
\tilde{\bm{\mathcal{A}}}_{\perp}(x),
\label{eq:eom_a}
\end{align}
where $\tilde{\bm{\mathcal{A}}}_{\perp}(x)$ is the transverse component of the gauge-invariant potential.  
Without loss of generality, we here consider a monochromatic incident field, ${\bm A}^{\rm ext}(x)={\bm A}^{\rm ext}({\bm x})e^{-i\Omega t}$. By performing the Fourier transformation on ${\bm x}$, one can solve Eqs.~\eqref{eq:eom_phase} and \eqref{eq:eom_a} as
\begin{gather}
\varphi({\bm x},\Omega) = \int \frac{d{\bm q}_{\perp}}{(2\pi)^2}e^{i{\bm q}_{\perp}\cdot{\bm x}}
\frac{{\bm q}_{\perp}\cdot{\bm A}^{\rm ext}_{\perp}({\bm q}_{\perp},\Omega)}{(\Omega/v)^2-q^2_{\perp}-(\omega_{\rm p}/v)^2}, \label{eq:phi2} \\
{\bm A}({\bm x},\Omega) = -\int \frac{d{\bm q}_{\perp}}{(2\pi)^2}e^{i{\bm q}_{\perp}\cdot{\bm x}}
\frac{{\bm A}^{\rm ext}_{\perp}({\bm q}_{\perp},\Omega)}{1+(vq_{\perp}/\omega_{\rm p})^2 - (\Omega/\omega_{\rm p})^2},\label{eq:a2}
\end{gather}
where ${\bm A}^{\rm ext}_{\perp}({\bm q}_{\perp},\Omega)\equiv u_{p,m}({\bm q}_{\perp})\hat{\bm e}_s$ is the transverse components of the external field and ${\bm q}_{\perp}=(q_x,q_y,0)$ is the momentum in the $x$-$y$ plane. The phase fluctuation is associated with the density fluctuation through the conservation law, i.e., the gauge invariance, which converts the phase mode into a plasmon with the frequency $\omega_{\rm p}= c/\lambda$ in charged systems. The THz frequency range, $\Omega \sim \Delta_0$, is far from the dispersion of the plasma mode, $\sqrt{\omega_{\rm p}^2+q^2_{\perp}v^2}$, and the phase mode is the off-resonant excitation. Therefore, the phase mode and charge density excitations are only transiently excited by the vortex beam and screened immediately after the pulsed vortex beam is turned off.\footnote{The effective field theory in Eq.~\eqref{eq:Seff} does not incorporate the polarization effect of the electrons, and the phase mode is gapped out to the plasma frequency of three-dimensional bulk systems. However, it has been shown that the plasma mode softens and may even become gapless in two-dimensional superconducting films or on the surface of three-dimensional superconductors~\cite{sun20,rad22}. Even though the dispersion of the plasma (phase) mode is typically impacted by the polarization effect, the key results presented in this work -- the amplification of the nonlinear Higgs excitation and the THG intensity through the off-resonant phase mode -- remain unchanged even when accounting for the polarization effect. In fact, the characteristics of the Higgs excitations and the THG intensity in two dimensions are similar to those in three dimensions, as shown in Sec.~S3.}

As shown in Eq.~\eqref{eq:ap}, the intensity $|{\bm A}^{\rm ext}({\bm q})|$ takes on a doughnut shape and peaks around $q_{\perp}\sim 1/w_0$. Since $q_{\perp} \ll \omega_{\rm p}/v$ for $w_0 \gg \xi_0$, the $q_{\perp}$ dependence can be removed from the denominator of Eq.~\eqref{eq:a2}. Thus, Eq.~\eqref{eq:a2} can be rewritten to 
\beq
\bm{ A}({\bm x},\Omega)  \approx 
- \frac{{\bm A}^{\rm ext}_{\perp}({\bm x},\Omega)}{1 - (\kappa_{\rm GL}/C)^2},
\eeq
where $C\equiv (c/v)(\Delta_0/\Omega)\sim c/v$ for $\Omega \sim \Delta_0$ corresponding to the nonlinear resonance to the Higgs excitation. Similarly, Eq.~\eqref{eq:phi2} can be simplified to $\varphi(x) \approx i\frac{v^2}{\omega^2_{\rm p}}{\bm \nabla}_{\perp}\cdot{\bm A}^{\rm ext}_{\perp}(x)$. This implies that the transverse components of the gauge-invariant potential are represented as
\begin{align}
\tilde{\bm{\mathcal{A}}}_{\perp}
= {\bm A}_{\perp} + {\bm A}^{\rm ext}_{\perp}
- \frac{c}{2e}{\bm \nabla}_{\perp}\varphi
\equiv {\bm{\mathcal{A}}}^{({\rm a})}_{\perp}
+ {\bm{\mathcal{A}}}^{({\rm b})}_{\perp}
\end{align}
and the longitudinal component is  
$\tilde{\mathcal{A}}_{z}= \hat{\bm z}A_z^{\rm ext}$.
The first term of the transverse components, $\bm{\mathcal{A}}^{({\rm a})}_{\perp}
\equiv {\bm A}_{\perp} + {\bm A}^{\rm ext}_{\perp}$, reads 
\begin{align}
\bm{\mathcal{A}}^{({\rm a})}_{\perp}({\bm x},\Omega)
\approx 
-\frac{(\kappa_{\rm GL}/C)^2}{1-(\kappa_{\rm GL}/C)^2}{\bm A}_{\perp}^{\rm ext}({\bm x},\Omega),
\label{eq:Aa} 
\end{align}
and the second term, ${\bm{\mathcal{A}}}^{({\rm b})}_{\perp}
\equiv - \frac{c}{2e}{\bm \nabla}_{\perp}\varphi$, is the contribution of the off-resonant plasma oscillation,
\begin{align}
{\bm{\mathcal{A}}}^{({\rm b})}_{\perp}({\bm x},\Omega)
=& -\frac{ic}{2e}\left(\frac{v}{\omega_{\rm p}}\right)^2
{\bm \nabla}_{\perp}\left({\bm \nabla}_{\perp}\cdot {\bm A}_{\perp}^{\rm ext}({\bm x},\Omega)\right).
\label{eq:Ab}
\end{align}
The transverse component of the gauge-invariant potential, $\tilde{\bm{\mathcal{A}}}^{({\rm a})}_{\perp}$, is influenced by the GL parameter, while remains unaffected by the OAM of light, $m$. The OAM dependency arises from the second term which involves the off-resonant phase oscillation with $\tilde{\bm{\mathcal{A}}}^{({\rm b})}_{\perp}\propto \hat{\bm e}_s\cdot{\bm \nabla}_{\perp}e^{im\theta}\sim m e^{iJ\theta}$. Hence, the contribution of the off-resonant phase excitation to $\tilde{\bm{\mathcal{A}}}$ is proportional to the OAM of light, $m$, and the spatial pattern reflects the total angular momentum, $J$.

Equation~\eqref{eq:Aa} indicates that in the low $\kappa_{\rm GL}$ regime, the internal field is induced to fully screen the transverse component of the external field as ${\bm A}_{\perp}({\bm x},\Omega)\approx - {\bm A}^{\rm ext}_{\perp}({\bm x},\Omega)$. As a consequence, the internal current generates a screening effect for the longitudinal component $B^{\rm ext}_z$, but the transverse components, $(B^{\rm ext}_x,B^{\rm ext}_y)$, remains unscreened when the thickness of the superconducting film is much smaller than $\lambda$. Thus, the total magnetic field in the superconducting film can be approximated as $\bm{\mathcal{B}}\approx (B^{\rm ext}_x,B^{\rm ext}_y,0)$ in the low $\kappa_{\rm GL}$ limit and $\bm{\mathcal{B}}\approx (B^{\rm ext}_x,B^{\rm ext}_y,B^{\rm ext}_z)$ in the type-II limit.

Let ${\bm j}^{(1)}$ be the linear current response to the vortex beam, which is obtained from Eq.~\eqref{eq:current} as 
\begin{align}
{\bm j}^{(1)}= - \frac{8e^2\kappa \Delta^2_0}{c}\left( \tilde{\bm{\mathcal{A}}}^{({\rm a})}_{\perp}
+ \tilde{\bm{\mathcal{A}}}^{({\rm b})}_{\perp}\right)
\equiv {\bm j}^{({\rm 1a})}+{\bm j}^{({\rm 1b})}.
\label{eq:j1}
\end{align}
For a monochromatic wave, the diamagnetic response reduces to 
\begin{align}
{\bm j}^{({\rm 1a})}({\bm x},t)
\propto \frac{u_{p,m}(\rho)}{1-(C/\kappa_{\rm GL}^2)}\left[
\cos(m\theta-\Omega t)\hat{\bm e}_x
- \sin(m\theta-\Omega t)\hat{\bm e}_y
\right],
\end{align}
and the off-resonant plasma oscillation reads
\begin{align}
{\bm j}^{({\rm 1b})}({\bm x},t) 
\propto&  \frac{m}{\rho}\left(\partial_{\rho}u_{p,m}(\rho)\right)\sin(J\theta-\Omega t)\hat{\bm e}_{\rho}\nn \\
&+ \frac{m^2}{\rho^2}u_{p,m}(\rho)\cos(J\theta-\Omega t)\hat{\bm e}_{\theta},
\end{align}
where $\hat{\bm e}_{\rho}$ and $\hat{\bm e}_{\theta}$ are the unit vectors along the radial and azimuthal directions, respectively. The first term in Eq.~\eqref{eq:j1} reflects the intensity of the Laguerre-Gaussian vortex beam in the shape of a doughnut, $|{\bm j}^{({\rm 1a})}|\propto|A_{\perp}^{\rm ext}({\bm r})|$, which peaks around $\rho \sim w_0$. On the other hand, the strength of the second term, denoted as $|{\bm j}^{({\rm 1b})}|$, is influenced by the OAM of light, $m$. The spatial oscillation pattern along the azimuthal direction reflects the total angular momentum $J=m+s$. For Gaussian beams with $m=0$, the off-resonant plasma oscillation is not responsible for the linear current response, resulting in ${\bm j}^{(1)}={\bm j}^{({\rm 1a})}$.

In the low $\kappa_{\rm GL}$ regime, the internal field is induced to screen the transverse component of the external field as $\bm{ A}({\bm x},\Omega) \approx - {\bm A}^{\rm ext}_{\perp}({\bm x},\Omega)$.
In this regime, the linear current response can be approximated as ${\bm j}({\bm x},\Omega)\approx {\bm j}^{({\rm 1b})}({\bm x},\Omega)\propto (\hat{\bm e}_s\cdot{\bm \nabla}_{\perp}){\bm \nabla}_{\perp}u_{p,m}({\bm x})$. The spatial pattern of the plasma (phase) oscillation encodes the OAM of light through $\varphi \propto {\bm \nabla}_{\perp}{\bm A}^{\rm ext} \propto m\cos(J\theta -\Omega t)$. Figure~\ref{fig:pulse}(e) displays ${\bm j}(x,y,t)$ for $(m,s)=(4,1)$ and $\Omega = 1.025\Delta_0$ at $\kappa_{\rm GL}=10$. For $\kappa_{\rm GL}=10$, the screening effect suppresses the first term of the current response, ${\bm j}^{({\rm 1a})}\propto \tilde{\bm{\mathcal{A}}}^{({\rm a})}_{\perp}$, and ${\bm j}^{({\rm 1b})}$ dominates the linear current response. Along the azimuthal direction in Fig.~\ref{fig:pulse}(e), the spatial profile coincides with the contribution of the off-resonant plasma oscillation, ${\bm j}^{({\rm 1b})}$. The oscillation pattern reflects the total angular momentum $J$, rather than the OAM $m$, where the phase mode plays an essential role on the transfer of the total angular momentum of light.
%

\begin{figure}[t!]
\includegraphics[width=85mm]{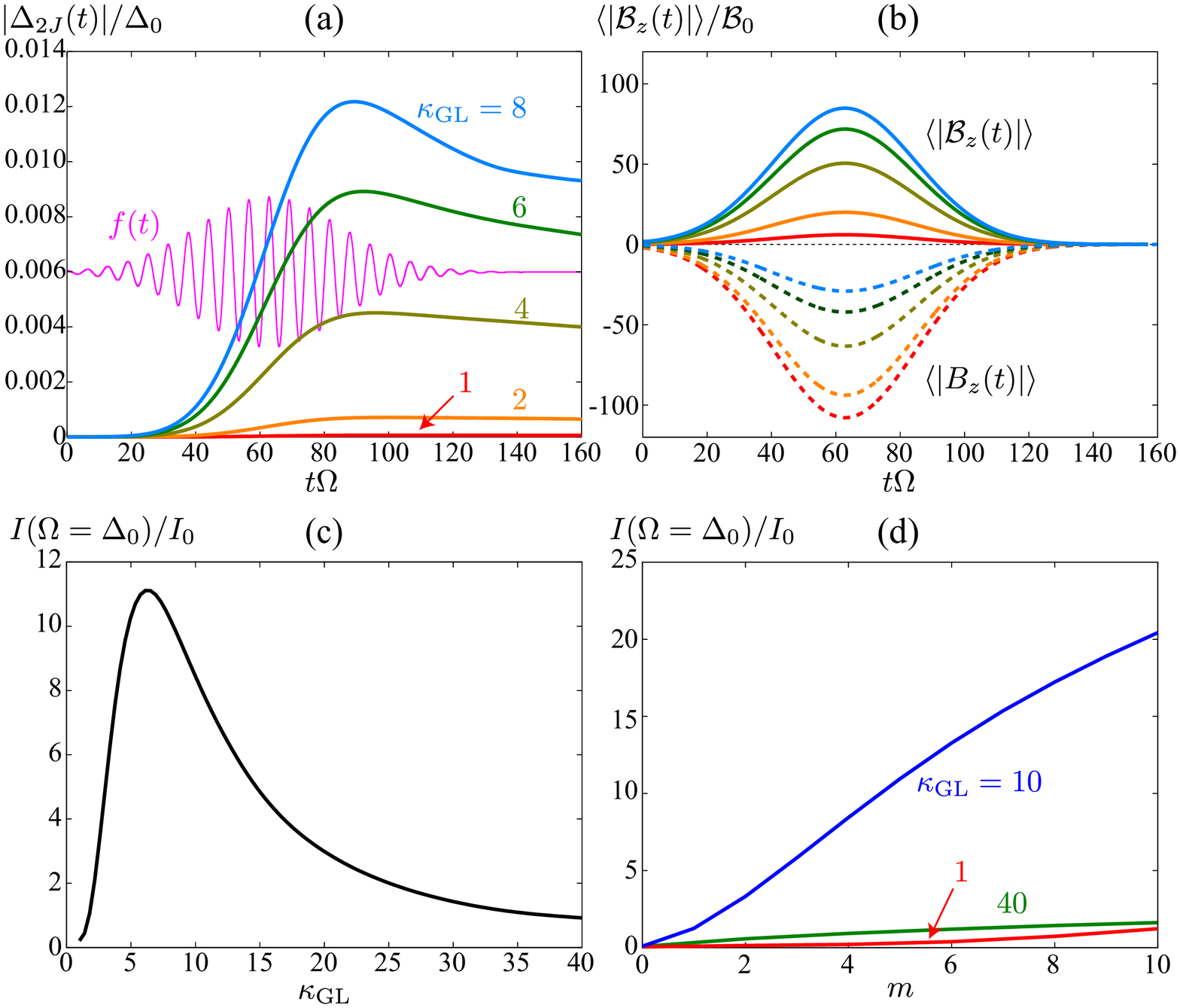}
\caption{(a) Angular momentum component $\Delta_{2J}(t)\equiv \int d{\bm x}e^{-i2J\theta}\Delta({\bm x},t)$ and (b) spatially averaged magnetic field $\langle \mathcal{B}_z(t)\rangle \equiv \int d{\bm x} |\mathcal{B}_z({\bm x},t)|$ (solid curves) and $\langle B_z(t)\rangle \equiv \int d{\bm x} |{B}_z({\bm x},t)|$ for $\kappa = 1$, 2, 4, where the total magnetic field is defined as $\bm{\mathcal{B}}\equiv {\bm \nabla}\times \bm{\mathcal{A}}={\bm \nabla}\times {\bm A}+{\bm B}_{\rm ext}$. Here we set $s=1$. (c) Third harmonic generation intensity $I_3(\Omega)/I_0$ as a function of $\kappa_{\rm GL}$, where we set $(m,s)=(4,1)$ and $\Omega = 1.025\Delta_0$ which is the nonlinear resonance to the Higgs mode. We have introduced $I_0\equiv c^2/(16\pi e)$. (d) $m$-dependence of $I_3(\Omega)/I_0$ at $\Omega = 1.025\Delta_0$ for $\kappa_{\rm GL}=1$, $10$, and $40$. We note that the THG intensity at $m=0$ is weak but finite.}
\label{fig:kappa}
\end{figure}

When the type-II limit ($\kappa_{\rm GL}\rightarrow \infty$) is approached, however, the screening current weakens and the external vector potential can penetrate the film, resulting in $\bm{ A}({\bm x},\Omega) \approx {\bm 0}$ and $\bm{\mathcal{A}}({\bm x},\Omega) \approx \bm{ A}^{\rm ext}({\bm x},\Omega) $. 
In this limit, the spatial pattern of the linear current response shows the OAM with ${\bm j}({\bm x},t) = -\frac{8e^2\kappa \Delta^2_0}{c}{\rm Re}{\bm A}_{\perp}^{\rm ext}({\bm x},t)\propto \cos(m\theta-\Omega t)\hat{\bm e}_s$. The spatial pattern is determined by the OAM of light, $m$, as the phase mode cannot be stimulated.

Let us now discuss the $\kappa_{\rm GL}$-dependencies of the nonlinear condensate and current responses. In Fig.~\ref{fig:kappa}(a), we present the angular momentum component of the condensate, $\Delta_{2J}(t)\equiv \int d{\bm x}e^{-i2J\theta}\Delta({\bm x},t)$, for $\kappa_{\rm GL}=1$, 2, 4, 6, and 8, where $(m,s)=(4,1)$ and $\Omega = 1.025\Delta_0$ are fixed. We find that the amplitude excitation is suppressed for low $\kappa_{\rm GL}$ values ($\kappa_{\rm GL}\le 1$), and the longitudinal component of the total magnetic field is fully screened, resulting in $\mathcal{B}_z \equiv ({\bm \nabla}\times {\bm A})_z+B^{\rm ext}_z \approx 0$. However, increasing $\kappa_{\rm GL}$ amplifies the nonlinear Higgs excitation, $|\Delta_{2J}(t)|$, where the longitudinal magnetic field cannot be fully screened. The THG intensity is enhanced as $\delta \Delta$ grows with $\kappa_{\rm GL}$. In Fig.~\ref{fig:kappa}(c), we plot the $\kappa_{\rm GL}$ dependence of the THG intensity $I_3(\Omega)/I_0$ for $(m,s)=(4,1)$ at $\Omega = 1.025\Delta_0$. We have introduced the constant $I_0\equiv c^2/(16\pi e)$. In the case of $(m,s)=(4,1)$, the optimal value is $\kappa_{\rm GL}\sim 10$. From Fig.~\ref{fig:kappa}(d), it is seen that the intensity of the THG significantly grows with the OAM of light for the optimal value of the GL parameter, $\kappa_{\rm GL}=10$. For $\kappa_{\rm GL}=40$, however, the THG intensity is insensitive to $m$. We note that the THG intensity at $m=0$ is weak but finite. Unlike vortex beams with $m\neq 0$, the Gaussian beam with $m=0$ involves only the transverse components of the EM fields. The Gaussian beam without the longitudinal component cannot drive the phase oscillation, resulting in the weak intensity of the THG.

The dependences of the Higgs excitations and the THG intensity on $\kappa_{\rm GL}$ and $m$ are attributed to the interplay between the screening effect and off-resonant plasma oscillation. To clarify this, we focus on the third-order current response via the nonlinear Higgs excitation,
\beq
{\bm j}^{(3)}({\bm x},t) = -\frac{16e^2\kappa\Delta_0}{c}\delta\Delta({\bm x},t)\tilde{\bm{\mathcal{A}}}({\bm x},t).
\eeq
The coupling of $\delta\Delta$ and $\tilde{\bm{\mathcal{A}}}({\bm x},t)$ is quadratic in the effective action, and ${\bm j}^{(3)}\propto\tilde{\bm{\mathcal{A}}}^2\tilde{\bm{\mathcal{A}}}$. As mentioned above, the gauge-invariant potential $\tilde{\bm{\mathcal{A}}}$ consists of the external vector potential and the off-resonant plasma excitation. The former is insensitive to the OAM of light and subject to the screening effect, while the intensity of the latter term depends on $m$ as mentioned in Eq.~\eqref{eq:Ab}. 
At the type-I limit ($\kappa_{\rm GL}\rightarrow 0$), the external field generated by the vortex beam is fully screened and the gauge-invariant potential becomes vanishingly small, resulting in $I_3\approx 0$. 
As $\kappa_{\rm GL}$ increases, the screening effect of the external field (${\bm A}^{\rm eff}$) weakens, and both the phase and amplitude of the condensate can experience the external potential, resulting in the amplification of the nonlinear Higgs response and the THG intensity with $\kappa_{\rm GL}$. As seen in Fig.~\ref{fig:kappa}(c), the THG intensity peaks around $\kappa_{\rm GL}\sim 10$.
As $\kappa_{\rm GL}$ further increases and approaches the type-II limit, however, the screening effect is absent and the gauge-invariant potential reduces to $\bm{\mathcal{A}}\approx {\bm A}^{\rm ext}$. Since the unscreened external potential ${\bm A}^{\rm ext}$ has a large intensity comparable to the electric field of $E_0=0.4~{\rm kV}/{\rm cm}$, it dominates the gauge-invariant coupling to the Higgs mode and the contribution of the off-resonant plasma excitation through ${\bm \nabla}\varphi$ becomes negligible, i.e., $\tilde{\bm{\mathcal{A}}}\approx {\bm A}^{\rm ext}$. Therefore, the absence of the screening effect at $\kappa_{\rm GL}\gg 1$ makes the THG intensity insensitive to the OAM of light. There exists an optimal value of $\kappa_{\rm GL}$ where the THG intensity becomes maximal. The optimal value depends on vortex beam parameters such as $(m,s)$, $\Omega$, and $w_0$.

\subsection{S2.5 Optimal beam waist for nonlinear Higgs excitations}

\begin{figure}[t!]
\includegraphics[width=85mm]{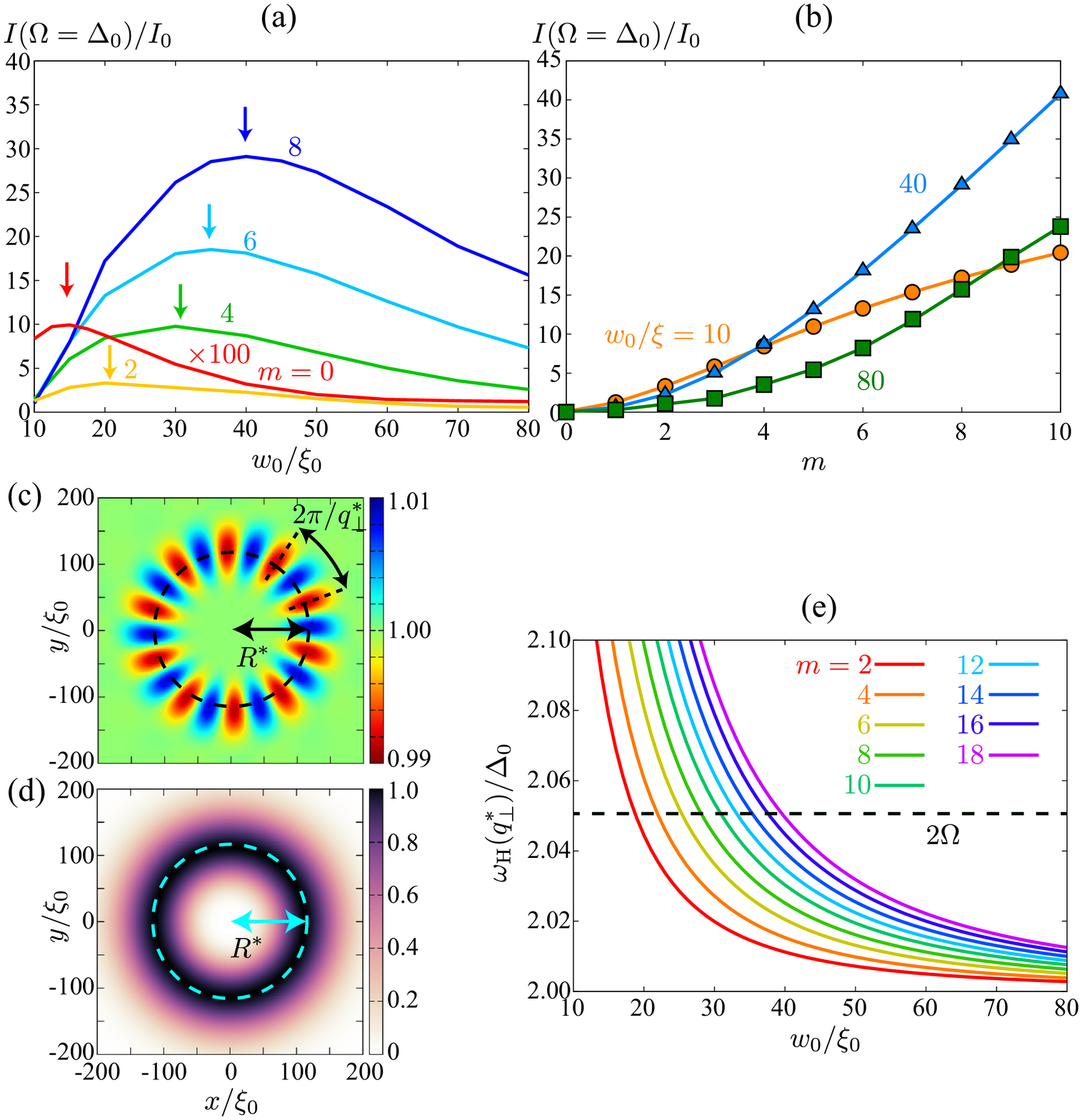}
\caption{(a) Beam waist ($w_0$) dependences of $I_3(\Omega)/I_0$ for $m=0$, 2, 4, 6, and 8, where we set $s=1$ and $\Omega = 1.025\Delta_0$. (b) $I_3(\Omega)/I_0$ as a function of $m$ for $w_0=10\xi_0$ (circles), $40\xi_0$ (triangles), and $80\xi_0$ (squares). We have introduced $I_0\equiv c^2/(16\pi e)$. (c,d) Snapshots of the order parameter amplitude $|\Delta(x,y,t)|$ (c) and the vortex beam intensity $|{\bm A}^{\rm ext}(x,y,t)|$ (d) for $w_0/\xi_0 = 80$, $(m,s)=(4,1)$, and $\Omega = 1.025\Delta_0$ at $t=153$. (e) The frequency of the resonant Higgs mode with the wave number $q_{\ast}\equiv 2J/R^{\ast}$, $\omega_{\rm H}(q^{\ast})$, as a function of $w_0$ for several $m$'s. }
\label{fig:waist}
\end{figure}

Lastly, we discuss the impact of the beam waist $w_0$. In Figs.~\ref{fig:waist}(a) and \ref{fig:waist}(b), we present the beam waist ($w_0$) dependences of $I_3(\Omega)/I_0$ for various $m$'s and the $m$ dependences for different $w/\xi_0$'s at $\Omega = 1.025\Delta_0$, respectively. As shown in Fig.~\ref{fig:waist}(a), there exists an optimal beam waist ($w_0^{\ast}$) at which the THG intensity is maximal, and $w_0^{\ast}$ increases as $m$ increases. The enhancement of the THG intensity around the optimal values $w_0^{\ast}$ can be attributed to the nonlinear resonance to the Higgs excitation. The gauge-invariant potential induced by the vortex beam leads to the nonlinear Higgs excitation at the frequency $2\Omega = \omega_{\rm H}(q_{\perp})$. The nonlinear response is expressed as 
\beq
\delta \Delta({\bm x},\Omega) = \int \frac{d{\bm q}_{\perp}}{(2\pi)^2} e^{i{\bm q}_{\perp}\cdot{\bm x}}\frac{\tilde{\bm{\mathcal{A}}}^2({\bm q}_{\perp},\Omega)}{(2\Omega)^2-(\omega_{\rm H}(q_{\perp}))^2},
\eeq
where $\omega_{\rm H}(q_{\perp}) = \sqrt{(2\Delta_0)^2+(vq_{\perp})^2}$ is the dispersion of the Higgs mode.
For a fixed $\Omega$, the nonlinear response becomes more prominent at the wave vector $q^{\rm res}_{\perp}$, where $2\Omega = \omega_{\rm H}(q^{\rm res}_{\perp})$. In Figs.~\ref{fig:waist}(c) and \ref{fig:waist}(d), the standing wave of the amplitude with the period of $2\pi/q^{\ast}_{\perp}=2\pi R^{\ast}/2J$ is induced around $\rho = R^{\ast}$. This occurs where the external potential, $|{\bm A}^{\rm ext}({\bm x},t)|$, is maximal. The circumference length $2\pi R^{\ast}$ is comparable with $2\pi w_0$ but slightly depends on the beam waist and OAM of light [see also Eq.~\eqref{eq:ap}]. It is observed that the nonlinear Higgs response is intensified when the circumference length $2\pi R^{\ast}$ is commensurate with the period of the standing wave of the resonant Higgs mode, $2\pi/q^{\rm res}_{\perp}$. To verify this, we plot the frequency of the resonant Higgs mode, $\omega_{\rm H}(q^{\ast}_{\perp})$, with the wave number $q^{\ast}_{\perp}\equiv 2J/R^{\ast}$, as a function of $w_0$ for different $m$ values in Fig.~\ref{fig:waist}(e). The optimal beam waist $w_0^{\ast}$, at which the THG intensity is maximum, approximately corresponds to $w_0$ which satisfies $\omega_{\rm H}(q^{\ast}_{\perp})=2\Omega$. It is seen from Fig.~\ref{fig:waist}(e) that for $m=2$, the nonlinear resonance to the Higgs mode occurs around $w_0/\xi_0=20$, which coincides with the optimal value $w_0^{\ast}$ denoted by the arrow in Fig.~\ref{fig:waist}(a). As depicted in Fig.~\ref{fig:waist}(e), the value of $w_0/\xi_0=20$ increases as $m$ increases. Moreover, it should be noted that the optimal beam waist is influenced by the frequency $\Omega$. The optimal values increase as $\Omega$ lowers and approaches $2\Delta_0$.

\begin{figure}[t!]
\includegraphics[width=85mm]{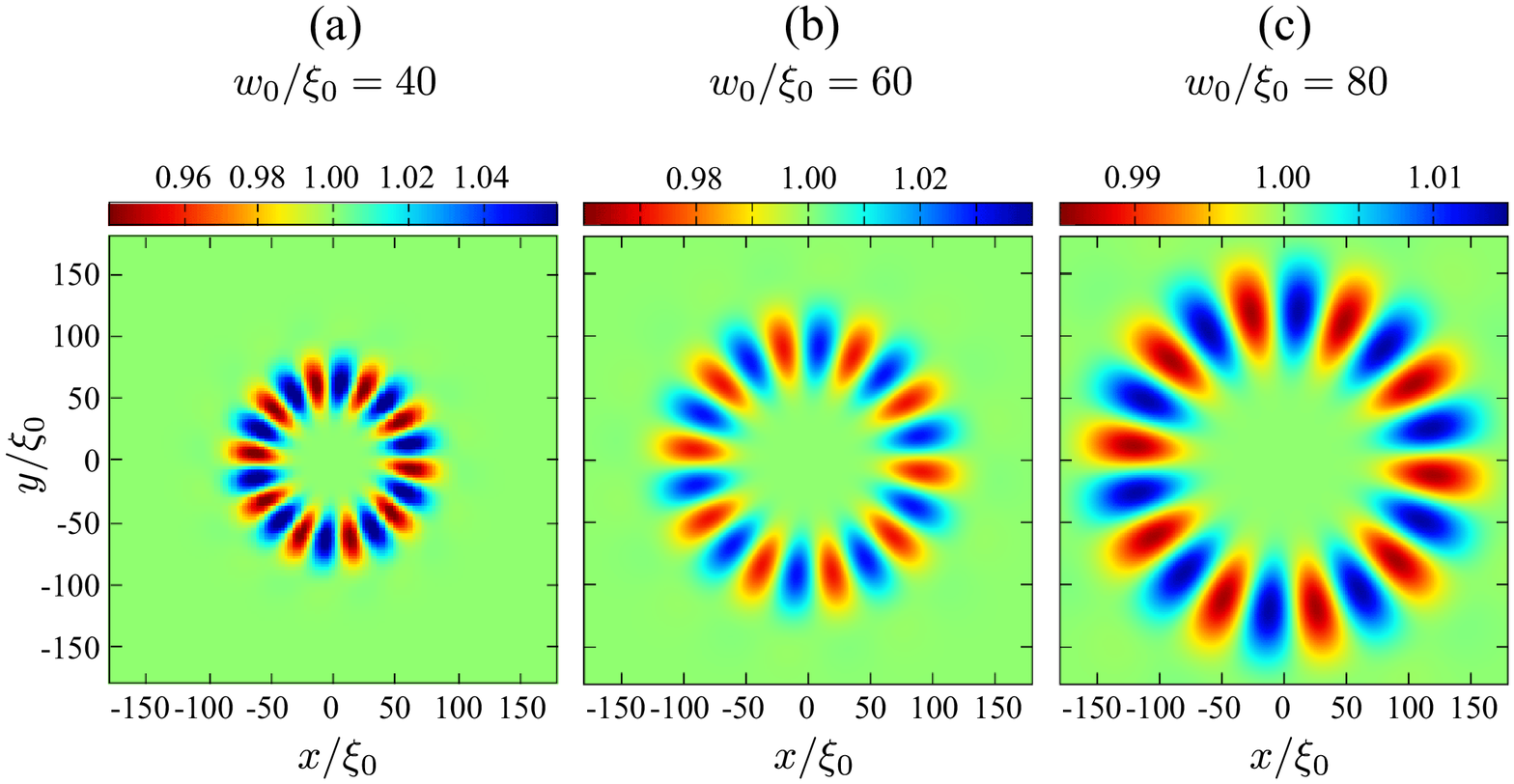}
\caption{Spatial profiles of the superconducting order parameter, $\Delta(x,y,t)$, at $t=85t_{\Delta}$ for $w_0/\xi_0=40$ (a), 60 (b), and 80 (c), where we set $(m,s)=(4,1)$ and $\Omega = 1.025\Delta_0$. In all data, we fix the intensity of ${\bm A}_{\rm ext}$ corresponding to $E_0=0.4~{\rm kV}/{\rm cm}$.}
\label{fig:waist2}
\end{figure}

\begin{figure*}[t!]
\includegraphics[width=180mm]{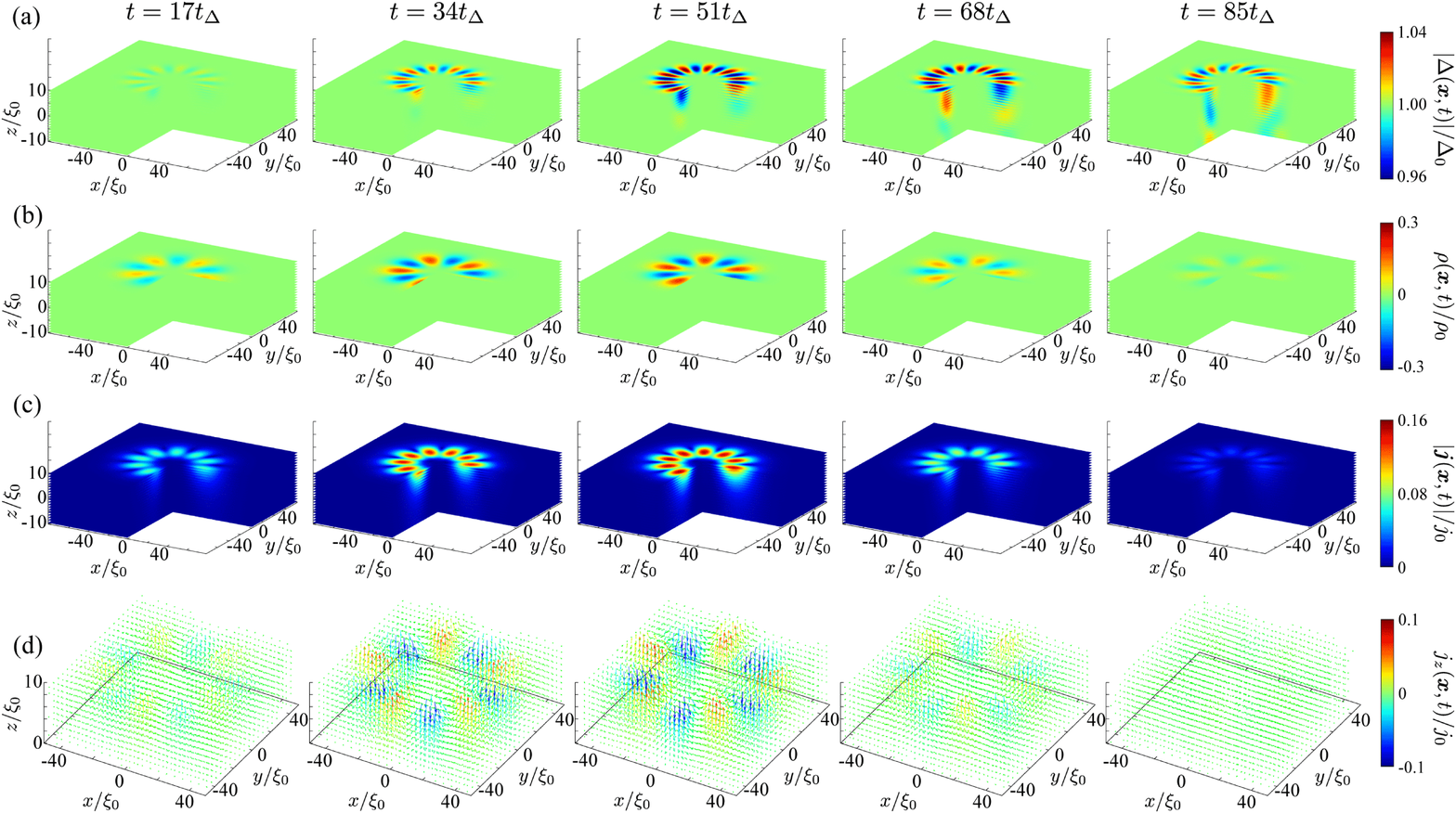}
\caption{Snapshots of $|\Delta({\bm x},t)|$ (a), $\rho({\bm x},t)$ (b), $|{\bm j}_{\parallel}({\bm x},t)|$ (c), and ${\bm j}({\bm x},t)$ (d), after the pulsed vortex beam with $(m,s)=(4,1)$ and $\Omega=1.025\Delta_0$ is irradiated at $t=0$. The GL parameter is $\kappa_{\rm GL}=10$. The other parameters of the vortex beam are $n_{\rm p}=5$ and $w_0=20\xi_0$. In the bottom panels, the color depicts $j_z({\bm x},t)/j_0$. The intensity of the pulsed vortex beam almost becomes maximum around $t=62t_{\Delta}$.}
\label{fig:snap}
\end{figure*}

Here we would also like to mention that in our numerical calculation, we used the beam waist $w_0$ that is smaller than the realistic values in THz frequency band. However, increasing the beam waist does not affect our main findings. In fact, as demonstrated in Fig.~\ref{fig:waist}(b), the enhancement of the THG intensity with the OAM of light is always observed for $w_0/\xi_0=10,40, 80$. Furthermore, the change in the beam waist does not alter the spatial pattern of the induced spiral Higgs wave (Fig.~\ref{fig:waist2}). The extent of the modulation in $\Delta(x,y,t)$ is proportional to $w_0/\xi_0$. The diffraction limit at THz frequency band imposes a size $w_0$ of $O(100\mu {\rm m})$, larger than the maximum size in our numerical simulation, $w_0/\xi_0=80$. However, realistic values of the beam waist imposed by the diffraction limit for a THz vortex beam can still lead to similar results for the spiral Higgs excitations and the enhancement of the THG intensity.

\section{S3. Spiral waves in three dimensions}
\label{sec:3d}

In the final section S3, we present the numerical results in full three dimensional films. Here we consider a superconducting film with a thickness of $d=20\xi_0$ and the penetration depth is fixed to $\lambda=10\xi_0$ ($\kappa_{\rm GL}=10$). The numerical simulation is performed on the discretized grids in the rectangular box with $x/\xi_0,y/\xi_0\in[-100,100]$ and $z/\xi_0\in[-10,10]$. The vortex beam is irradiated towards the upper surface of the film at $z=d/2$. The irradiated vortex beam is implemented as the boundary condition on the upper surface as described in Sec.~S1.3, where the value of the internal magnetic field matches a magnetic field generated by the vortex beam, ${\bm B}^{\rm ext}$ in Eq.~\eqref{eq:bext}. 

Figure~\ref{fig:snap} shows the snapshots of $|\Delta({\bm x},t)|$, $\rho({\bm x},t)$, and $|{\bm j}_{\parallel}({\bm x},t)|=\sqrt{j^2_x({\bm x},t)+j^2_y({\bm x},t)}$, and the vectorial plot of ${\bm j}({\bm x},t)$ (from top to bottom), after the irradiation of the vortex beam with $(m,s)=(4,1)$. Here we set $\Omega=1.025\Delta_0$, $w_0=20\xi_0$, and $n_{\rm p}=5$. The numerical results are essentially the same as those observed in the case of two-dimensional simulations. Figures~\ref{fig:snap}(a) and \ref{fig:snap}(b) show that the vortex beam induces spiral waves in both the condensate amplitude and the phase excitation, respectively. The spatial profiles of the amplitude and phase excitations reflect twice the total angular momentum $2J$ and the total angular momentum $J$, respectively. The amplitude oscillation penetrates the skin depth within $\lambda=20$, while alternative positive and negative charge distributions accumulate on the surface as $\rho({\bm x},t)\sim A\sin(J\theta-\Omega t)\delta(z-d/2)$. Such charge distribution on the upper surface creates a three-dimensional flow of the supercurrent density, depicted in Fig.~\ref{fig:snap}(c,d). Two-dimensional simulations can accurately represent the nonlinear dynamics of superconductors induced by vortex beams. These essential features remain unchanged regardless of the dimensionality.

\bibliography{LGbeam}